\documentclass[reqno]{amsart}

\usepackage{amsmath}
\usepackage{amsthm}
\usepackage{amsfonts}

\newtheorem{thm}{Theorem}[section]
\newtheorem{lem}[thm]{Lemma}

\newtheorem{rem}[thm]{Remark}

\newcommand{\R}{\mathbb{R}}

\newcommand{\N}{\mathbb{N}}

\newcommand{\M}{\mathcal{M}}

\newcommand{\Om}{\Omega}
\newcommand{\vr}{\varrho}

\newcommand{\rd}{\mathrm{d}}
\newcommand{\divv}{\mathrm{div}}

\newcommand{\Wqb}{W_{p,\mathcal{B}}}

\newcommand{\bqn}{\begin{equation}}
\newcommand{\eqn}{\end{equation}}
\newcommand{\bqnn}{\begin{equation*}}
\newcommand{\eqnn}{\end{equation*}}
\newcommand{\bs}{\begin{split}}
\newcommand{\es}{\end{split}}
\newcommand{\bean}{\begin{eqnarray*}}
\newcommand{\eean}{\end{eqnarray*}}

\title[Proteus Mirabilis Swarm-Colony Development]
{On an Age and Spatially Structured Population Model for {\it
Proteus Mirabilis} Swarm-Colony Development}

\author[Ph. Lauren\c{c}ot and Ch. Walker]
{Philippe Lauren\c{c}ot and Christoph Walker}

\address{Institut de Math\'{e}matiques de Toulouse \\ CNRS (UMR 5219) \&
Universit\'{e} de Toulouse\\
118 route de Narbonne
\\ F--31062 Toulouse cedex 9\\
 France.}
\email{laurenco@mip.ups-tlse.fr}

\address{
Gottfried Wilhelm Leibniz Universit\"at Hannover\\
Institut f\"ur Angewandte Mathematik\\
Welfengarten 1\\ D--30167 Hannover}
\email{walker@ifam.uni-hannover.de}

\begin{document}

\begin{abstract}

{\it Proteus mirabilis} are bacteria that make strikingly regular
spatial-temporal patterns on agar surfaces. In this paper we
investigate a mathematical model that has been shown to display
these structures when solved numerically. The model consists of an
ordinary differential equation coupled with a partial differential
equation involving a first-order hyperbolic aging term together
with nonlinear degenerate diffusion. The system is shown to admit
global weak solutions.
\end{abstract}


\maketitle

\begin{center}
{\it Dedicated to Glenn F. Webb\\
on the occasion of his 65th birthday.}
\end{center}

\section{Introduction}

Bacteria of the species {\it Proteus mirabilis} are ubiquitous
throughout nature. In human beings, {\it Proteus mirabilis} is
found as part of the normal flora of the gut. Its main
pathological role is in infections of the urinary tract, but it
can also cause wound infections and {\it septicaemia}. Even though
most of the human urinary tract infections are due to the
bacterium {\it Escherichia coli}, urinary tract infections due to
{\it Proteus mirabilis} are also well-documented. It commonly
invades the urinary tract when the normal function of the tract is
disturbed by instrumentation such as catheterization. Once
attached to urinary tract, {\it Proteus mirabilis} infects the
kidney more commonly than {\it Escherichia coli} and
characteristically leads to urinary stones.

{\it Proteus mirabilis} can exist in two distinct morphological
and physiological forms known as ``swimmer" cells and ``swarmer"
cells, respectively. Broth cultures of {\it Proteus mirabilis}
consist virtually exclusively of mononuclear cells (swimmers)
approximately 1 $\mu m$ wide with short flagella. Swimmer cells go
through a prototypical bacterial cell growth and division cycle.
However, when inoculated onto agar surfaces, some cells cease
septation but continue to grow and produce many lateral flagella
to form elongated multi-nucleoid hyperflagellated swarmer cells up
to 100 $\mu m$ in length which aggregate in parallel arrays to
form motile multicellular ``rafts". The process in which dividing
cells become swarmers is called ``differentiation" and occurs only
above a critical dividing-cell density. Rafts of swarmer cells are
capable of translocation while swimmer cells are immobile. The
movement of {\it Proteus mirabilis} through raft building requires
two things, namely sufficient maturity in swarmer cells to
contribute to raft building and sufficient biomass of mature cells
to form the rafts. After some time migrating, when the
multinuclear swarmer cells approach a maximal size, they cease
movement and rapidly ``dedifferentiate" again into single nucleus
swimmer cells. This coordinated burst of swarming activity
interspersed with a consolidation to the swimmer state results in
characteristic concentric rings of growth. It is due to these
strikingly regular spatial and temporal patterns that {\it Proteus
mirabilis} has attracted attention in the mathematical biology
literature \cite{Ayati1,ES,MKK}.

The ability to form swarmer cells seems to allow rapid
colonization of solid surfaces and the establishment of extensive
{\it Proteus mirabilis} biofilms. Apparently, {\it Proteus
mirabilis} can also swarm over the surfaces of all the major
catheter. Swarming may thus play roles in both the initiation of
catheter associated infections and the subsequent spread of the
biofilm over the catheter surface.

The processes involved in the evolution of {\it Proteus mirabilis}
and the formation of regular patterns are rather complex. A key
ingredient of the mathematical representation is the age
dependence of swarmer cell behavior. An age and spatially
structured model for {\it Proteus mirabilis} swarm colony
development was presented in \cite{ES}, and - in slightly modified
form - in \cite{MKK} and \cite{Ayati1,Ayati2}. Denoting by
$v=v(t,x)$ the swimmer cell density in dependence of time $t\ge 0$
and spatial position $x\in \Om$ (with a spatial region
$\Om\subset\R^n$) and by $u=u(t,a,x)$ the swarmer cell density
which additionally depends on age $a\ge 0$, the models in
\cite{Ayati1,Ayati2,ES,MKK} can be re-cast in the form
\begin{align}
\partial_t u+\partial_a u&\,=\, \divv_x
\big(D(\Lambda(t,x))\,\nabla_x u\big)\, -\,\mu(a)\, u\ , \quad
(t,a,x)\in
(0,\infty)\times(0,\infty)\times\Om\ ,\label{1}\\
\partial_t v&\,= \frac{1}{\tau}\,\big(1\,-\,\xi(v)\big)\, v\,+\int_0^\infty
e^{a/\tau}\,\mu(a)\, u(t,a,x)\, \rd a\ ,  \quad (t,x)\in
(0,\infty)\times\Om\
,\label{2}\\
u(t,0,x)&\,=\, \frac{1}{\tau}\,\xi\big(v(t,x)\big)\,v(t,x)\ ,
\quad
(t,x)\in(0,\infty)\times\Om\ ,\label{3}\\
u(0,a,x)&\,=\,u^0(a,x)\ ,\qquad v(0,x)=v^0(x)\ , \qquad (a,x)\in
(0,\infty)\times\Om \ ,\label{4}\\
\partial_\nu u&\,=\,0\ ,\quad (t,a,x)\in
   (0,\infty)\times (0,\infty)\times \partial\Om\ ,\label{5}
\end{align}
where
    \bqn\label{6}
    \Lambda(t,x)\,:=\int_{a_0}^\infty e^{a/\tau}\, u(t,a,x)\, \rd a\
,\quad (t,x)\in (0,\infty)\times\Om\ .
    \eqn
The major differences of the models \cite{Ayati1,ES,MKK} and their
philosophies are in different choices of the functions $D$, $\mu$,
and $\xi$. The meaning of the various terms are as follows:
$\Lambda=\Lambda(t,x)$ represents the total motile swarmer cell
biomass, where $a_{0}\ge 0$ is the minimal age of swarmer cells
required to participate actively in group migration. The
exponential comes in since biomass increase during swarm
development is assumed to occur at the same rate as during the
swimmer cell cycle. The parameter $\tau$ is the time it takes a
cell to subdivide.

Equation \eqref{1} expresses the change in time of swarmer cells
of a given age $a$. Movement of {\it Proteus mirabilis} occurs if
sufficiently many swarmers above the critical age $a_{0}$ group
together to build a mass above a certain threshold
$\Lambda_{min}\ge 0$. Thus, the diffusivity $D$ depends on
$\Lambda$ and is small (or zero) for $\Lambda$ small. For
instance, $D$ may be of the form
    \bqn\label{8}
    D(\Lambda)\,= \,D_0\, \max\{\Lambda-\Lambda_{min},0\}^{m-1}
    \eqn
as in \cite{Ayati1} with $D_0\in (0,\infty)$ and $m=2$. In
\cite{ES} also a dependence of $D$ on $v$ and on a memory term is
included, something we will refrain of taking into account. Note
that the exponential weighting in $\Lambda$ means in \eqref{8}
that older cells contribute more to swarming than younger cells.
The age dependent function $\mu$ in
\eqref{1} is the dedifferentiation modulus, which is higher for
older swarmers than for younger ones. A typical shape for $\mu$ is
a narrow hump located around a maximal age $a_{max}$ and zero
elsewhere. The limit choice $\mu(a)=\mu_0\, \delta_{a=a_{max}}$
has also been considered in \cite{ES}.

The change in time of the swimmer population is given by equation
\eqref{2}. The population grows exponentially with rate $1/\tau$.
Some of the swimmer cells cease septation and differentiate with
rate $\xi(v)/\tau$ into swarmers of age 0. This increase in
swarmer cells is reflected by equation \eqref{3}. As pointed out
in \cite{Ayati1} the function $\xi$ should be zero for $v$ small.
Indeed, the incorporation of a lag phase in swarmer cell
production triggers the development of a consolidation phase after
a swarm phase and thus prevents a self-sustaining soliton caused
by swarmers that dedifferentiate into swimmers immediately
differentiating into new swarmers. This lag in the onset of
differentiation was observed in \cite{WilliamsSchwarzhoff} and
included in the models in \cite{Ayati1,Ayati2,MKK}. The integral
term in \eqref{2} represents dedifferentiation of swarmer cells
into swimmer cells.

The basis of equations \eqref{1}-\eqref{5} was presented in
\cite{ES} and extensions and modifications of these equations were
proposed in \cite{Ayati1,Ayati2,MKK}. In \cite{ES} and
\cite{Ayati1,Ayati2} the main focus - besides the modeling
aspect~- were computational results displaying the spatial and
temporal patterns of concentric rings with equal width. In
\cite{Ayati2} numerical results were presented examining the
necessity of a sharp age of dedifferentiation from swarmer to
swimmer cells. All papers \cite{Ayati1,Ayati2,ES} use explicit age
dependence in the evolution of the swarmers. As pointed out in
\cite{Ayati1} explicit age structure provides a mechanism for
controlling - at least numerically - the ratio of time spent
swarming to time spent in consolidation without changing the total
cycle time. In \cite{MKK} a reaction-diffusion model for {\it
Proteus mirabilis} swarm-colony development based on averaging
over the age variable was used and results on the long time
distribution $\Lambda/v$ were derived. A model for the periodic
swarming of {\it Proteus} ignoring the age structure from the
outset was introduced in \cite{CMV}.

For further reading concerning morphology and pathogenicity of
{\it Proteus mirabilis} and for numerical results for models of
this bacteria we refer to
\cite{Ayati1,Ayati2,CMV,ES,MKK,MB,RaurpichEtAl,WCB,WilliamsSchwarzhoff}
and the references therein.

On the other hand, fewer mathematical results seem to be available
for models of {\it Proteus mirabilis} and the only result
regarding the mathematical well-posedness of models for {\it
Proteus mirabilis} we are aware of is \cite{Frenod}. Existence and
uniqueness of weak solutions to \eqref{1}-\eqref{5} are shown in
\cite{Frenod} for the case of non-degenerate diffusion including
memory.

The purpose of this paper is to prove an existence result for
diffusion coefficients $D(\Lambda)$ that may degenerate for
$\Lambda=0$ and thus to get closer to the biological reality.
Still we cannot handle the case where $D$ is given by \eqref{8}
but expect that the outcome of the model with diffusivity
$$
D(\Lambda)\,= \,D_0\, \max\{\Lambda-\Lambda_{min},0\}^{m-1} +
e^{-1/(\varepsilon\,\Lambda)}
$$
(to which our result applies if $m\ge 3$) for small
$\varepsilon>0$ resembles that for $\varepsilon=0$ from a
numerical viewpoint. However, proving the formation of regular
spatio-temporal patterns is beyond the scope of this paper.

The outline of the paper is as follows: In the next section we
first establish an existence and uniqueness result for the
non-degenerate case; that is, when $D$ is bounded below by a
positive constant. Our method for proving this result is
completely different from that in \cite{Frenod}. Section
\ref{sect3} then shows how to handle certain degenerate
diffusivity.

Throughout the paper we assume that the minimal age $a_0$ required
for swarmer cells to participate actively in the collective motion
is positive. The case $a_0=0$ turns out to be easier and could
also be handled with minor modifications.

\section{The Non-Degenerate Case}\label{sect2}

Throughout this section we suppose that the diffusivity $D$
satisfies
    \bqn\label{10}
    D\in C^{2-}(\R)\qquad\text{and}\qquad D(z)\ge
    d_0>0\quad\text{for}\quad z\in \R\ ,
    \eqn
where $C^{k-}$ (resp. $C_b^{k-}$) for $k\in \N\setminus\{0\}$
denotes the set of $C^{k-1}$-smooth functions with a Lipschitz
continuous (resp. uniformly Lipschitz continuous on bounded
subsets) $(k-1)$-th derivative. As for the differen\-tiation rate
$\xi$ we assume that
    \bqn\label{11}
    \xi\in C^{3-}(\R)\qquad\text{and}\qquad 0\le
    \xi(z)\le1\quad\text{for}\quad z\in \R\ ,
    \eqn
while the dedifferentiation modulus
    \bqn\label{12}
    \mu\in BC(\R):=C(\R)\cap L_\infty(\R)\quad\text{is non-negative}\ .
    \eqn
Let $\Om$ be a bounded and smooth domain in $\R^n$. We fix $p>n$
and denote by $\Wqb^{2\sigma}$ either the space
$W_{p}^{2\sigma}:=W_{p}^{2\sigma}(\Om)$ if $2\sigma\le 1+1/p$ or
the subspace of $W_{p}^{2\sigma}(\Om)$ consisting of those
elements satisfying homogeneous Neumann boundary conditions if
$2\sigma>1+1/p$. For abbreviation we put
$E_\sigma:=L_1(\R^+,\Wqb^{2\sigma},e^{a/\tau}\rd a)$ for
$\sigma\in [0,1]$.

In the following we denote by $c(T)$, $c(R)$, and $c(T,R)$
constants depending increasingly on the arguments and that may
differ from occurrence to occurrence .

We first prove an auxiliary result regarding the solvability of
\eqref{2}.

\begin{lem}\label{L1}
Let $T>0$, $2\sigma\in (1+n/p,2)$, and assume that $v^0\in\Wqb^2$
and $u\in C([0,T],E_\sigma)$ are given non-negative functions.
Then there exists a unique solution $v=v_u\in
C^1([0,T],\Wqb^{2\sigma})$ to \eqref{2} subject to the initial
condition $v(0)=v^0$. This solution is non-negative and belongs to
$C([0,T],\Wqb^2)$ if $u\in L_1([0,T],E_1)$.

Moreover, if $u$ and $\bar{u}$ both belong to $C([0,T],E_\sigma)$
and satisfy $\max{\left\{ \|u(t)\|_{E_\sigma}\,,\,
\|\bar{u}(t)\|_{E_\sigma} \right\}}\le R$ for $t\in [0,T]$ and
some $R>0$, then
    \bqn\label{14}
    \|v_u(t)-v_u(s)\|_{W_p^{2\sigma}}\le c(T,R)\,\vert t-s\vert\
    ,\quad 0\le t,s\le T\,
    \eqn
and
 \bqn\label{13A}
    \|v_u(t)-v_{\bar{u}}(t)\|_{W_p^{2\sigma}}\le
    c(T,R)\,\|u-\bar{u}\|_{C([0,T],E_\sigma)}
    ,\quad 0\le t\le T\,
    \eqn
for some constant $c(T,R)>0$.
\end{lem}

\begin{proof}
First note that the regularity of $\xi$ and
\cite[Theorem~4.2]{AmannMult} imply
    \bqn\label{13}
    [v\mapsto \xi(v)v ]\in C_b^{1-}(\Wqb^{2\sigma},\Wqb^{2\sigma})
    \eqn
since $2\sigma>1+n/p$. In addition, \eqref{12} ensures that the
integral term in \eqref{2} belongs to $C([0,T],\Wqb^{2\sigma})$.
The existence of a unique non-negative solution $v_u\in
C^1(J,\Wqb^{2\sigma})$ is now obvious, where either $J=[0,T]$ or
$J=[0,\tilde{T})$ with $\tilde{T}<T$ and
$\|v_u(t)\|_{W_p^{2\sigma}}\rightarrow\infty$ as $t\nearrow
\tilde{T}$. Next, \eqref{11} and the embedding
$W_p^{2\sigma}\hookrightarrow L_\infty$ ensure $v_u\in L_\infty
(J,L_\infty)$. Then, taking the gradient with respect to $x$ on
both sides of \eqref{2} we similarly obtain $v_u\in
L_\infty(J,W_\infty^1)$. Recalling that pointwise multiplication
satisfies
$$
W_p^{2\sigma-1}\times W_\infty^1\hookrightarrow
W_p^{2\sigma-1}\times W_p^{2\sigma-1}\hookrightarrow
W_p^{2\sigma-1}
$$
according to \cite[Theorem~4.2]{AmannMult} since $2\sigma-1>n/p$
we deduce that
$$
\|\xi(v_u(t))\, v_u(t)\|_{W_p^{2\sigma}}\,\le\,
c\big(1+\|v_u(t)\|_{W_p^{2\sigma}}\big)\ ,\quad t\in J\ .
$$
>From this we first conclude that $v_u\in
L_\infty(J,W_p^{2\sigma})$, whence $J=[0,T]$, and then $\xi(v)v\in
L_\infty (J,W_p^{2\sigma})$ so that \eqref{14} follows by
\eqref{2}. Property \eqref{13A} is implied by \eqref{12} and
\eqref{13}. Finally, if $u\in L_1([0,T],E_1)$, then the integral
term in \eqref{2} belongs to $L_1([0,T], \Wqb^2)$ due to
\eqref{12}, and we readily infer that $v_u$ belongs to
$C([0,T],\Wqb^2)$.
\end{proof}

The solvability of \eqref{1} is based on the following formal
observation: Suppose that the function $u$ is sufficiently smooth
so that the function $\Lambda=\Lambda_u$, given by \eqref{6},
leads to a well-defined evolution system  $U_{A_u}(t,s)$ on $L_p$
corresponding to the differential operator
    \bqn\label{200}
A_u(t)\, w:=-\divv_x\big(D(\Lambda_u(t))\nabla_x w\big)\ ,\quad
w\in \Wqb^2\ .
    \eqn
Then \eqref{1}, \eqref{3}-\eqref{5} can be re-written as a problem
in $L_p$ of the form
     \bqnn
    \begin{array}{rrlll}
    &\partial_t u\,+\,\partial_a u \,+\,\mu(a) u &=& -A_u(t) u\ , &
    a>0\ ,\  0<t\le T\ ,\\
    &u(t,0,\cdot) &=& \displaystyle{\frac{1}{\tau}\,
    \xi\big(v_u(t)\big)\, v_u(t)}\ ,& 0<t\le T\ , \\
    &u(0,a,\cdot)&=& u^0(a,\cdot)\ ,& a>0\ ,
    \end{array}
    \eqnn
where $v_u$ is the corresponding solution to \eqref{2}. Applying
the method of characteristics we derive that $u$ is a fixed point
of the map $\Phi$, given by
    \bqn\label{16A}
    \Phi(u)(t,a)\,:=\,\left\{\begin{array}{ll}
    \displaystyle{\frac{1}{\tau}\,e^{-\int_0^a\mu(r)\rd r}\,
    U_{A_u}(t,t-a)\,\xi(v_u(t-a))\,v_u(t-a)}\ ,& 0\le a\le t\ ,\\
    & \\
    \displaystyle{e^{-\int_{a-t}^a\mu(r)\rd r}\,
    U_{A_u}(t,0)\,u^0(a-t)}\ ,& 0\le t<a\ .
    \end{array}
    \right.
    \eqn
We now show that this map $\Phi$ indeed has a fixed point in a
suitable space and thus \eqref{1}-\eqref{5} admits a unique
solution. More precisely, we have:

\begin{thm}\label{T2}
Suppose \eqref{10}-\eqref{12} and fix $p>n$, $q>1$ and $2\omega\in
(1+n/p,2)$. Consider non-negative initial values $v^0\in \Wqb^2$
and
$$
u^0\in E_1\cap C^1([0,\infty),L_p)\cap
L_q\big((0,a_0),W_p^{2\omega}\big)\cap C([0,a_0],W_p^{2\omega-1})
$$
satisfying the compatibility condition $\xi(v^0)\,v^0=\tau\,u^0(0,
\cdot)$ in $\Om$. Then problem \eqref{1}-\eqref{5} possesses a
unique non-negative solution $(v,u)$ with
$$
    v\in C^1([0,\infty),\Wqb^{2\eta})\cap C([0,\infty),\Wqb^2)\ ,\quad
    u\in C([0,\infty), E_\eta)\cap L_{\infty, loc}([0,\infty), E_1)
$$
for any $\eta\in (0,1)$ and such that $u$ satisfies
    \begin{align*}
    \partial_t u(t,\cdot)\,,\, \partial_a u(t,\cdot)\in
    C([0,t],L_p)\cap C((t,\infty),L_p)\ ,\\
    \partial_t u(\cdot,a)\,,\, \partial_a u(\cdot,a)\in
    C([0,a),L_p)\cap C([a,\infty),L_p)
    \end{align*}
for all $t,a>0$ and solves \eqref{1} in $L_p$ for $t\not= a$.
\end{thm}

\begin{proof}
Given $\eta\in (0,1)$ we fix numbers $\vartheta, \sigma$, and
$\vr$ such that
$$1+n/p<2\vartheta<2\omega<2\sigma<2\vr<2$$ and $\eta<\sigma$ and choose
$\kappa\in (0,\min\{\sigma-\vartheta,1/q'\})$, where $q'$ is the
dual exponent of $q$. Note that we may assume without loss of
generality that $q\omega<1$ by making $q$ smaller if necessary.
Let $c_0$ be the norm of the natural injection
$\Wqb^{2\vr}\hookrightarrow \Wqb^{2\sigma}$ and let $R>0$ be such
that
    \bqn\label{2000}
    c_0\,e^{1/\tau}\,\|u^0\|_{E_\vr} +\|u^0\|_{L_q((0,a_0),W_p^{2\omega})}\,\le\, R\ .
    \eqn
For $T\in(0,1)$ we denote by $\mathcal{V}_T$ the space consisting
of all non-negative $u\in C([0,T],E_\sigma)$ such that
$\|u(t)\|_{E_\sigma}\le R+1$ and
$\|\Lambda_u(t)-\Lambda_u(s)\|_{W_p^{2\vartheta}}\le \vert
t-s\vert^\kappa$ for $0\le t,s\le T$, where $\Lambda_u$ is given
by \eqref{6}. Then, given any $u\in \mathcal{V}_T$, it follows
that the operator $-A_u(t)$ defined in \eqref{200} is for each
$t\in [0,T]$ the generator of a positive analytic semigroup on
$L_p$ (e.g. \cite{AmannIsrael,Tanabe}). Moreover, due to the
embedding $W_p^{2\sigma}\hookrightarrow
W_p^{2\vartheta}\hookrightarrow W_\infty^1$ we have
$$
\|A_u(t)-A_u(s)\|_{\mathcal{L}(\Wqb^2,L_p)}\,\le\, c(R)\, \vert
t-s\vert^\kappa\ ,\qquad 0\le t,s\le T\ ,
$$
and
    \bqn\label{15}
    \| A_u(t)-A_{\bar{u}}(t)\|_{\mathcal{L}(\Wqb^2,L_p)}\,\le\,
    c(R)\,\|u-\bar{u}\|_{\mathcal{V}_T}\ ,\qquad 0\le t\le  T\
    ,\quad u, \bar{u}\in \mathcal{V}_T\ ,
    \eqn
with the notation
$$
\|u\|_{\mathcal{V}_T}:= \sup_{t\in [0,T]}{\|u(t)\|_{E_\sigma}}
\;\;\mbox{ for }\;\; u\in \mathcal{V}_T\ .
$$
Therefore, invoking Corollary~II.4.4.2, Lemma~II.5.1.3,
Lemma~II.5.1.4, Equation~(II.5.3.8), and Section~II.6.4 in
\cite{LQPP} and using standard interpolation results on Sobolev
spaces with boundary conditions we derive that, for any $u\in
\mathcal{V}_T$, there exists a unique positive evolution system
$U_{A_u}(t,s)$, $0\le s\le t\le T$ on $L_p$ such that
    \bqn\label{16}
    \|U_{A_u}(t,s)\|_{\mathcal{L}(\Wqb^{2\alpha})}\,+\,
    (t-s)^{\gamma-\alpha}\|U_{A_u}(t,s)\|_{\mathcal{L}(\Wqb^{2\beta},
    \Wqb^{2\gamma})}\,\le\, c(R)
    \eqn
for $0\le s<t\le T$, $0\le \alpha <\beta\le \gamma\le 1$ with
$2\beta, 2\gamma \not= 1+1/p$, and
     \bqn\label{18}
    \|U_{A_u}(t,r)-U_{A_u}(s,r)\|_{\mathcal{L}(\Wqb^{2\gamma},\Wqb^{2\beta})}\,
    \le\,  c(R)\, (t-s)^{\gamma-\beta}
    \eqn
for $0\le r<s<t\le T$, $0< \beta \le \gamma < 1$ with $2\beta,
2\gamma \not= 1+1/p$. In addition, if $\bar{u}$ is another
function in $\mathcal{V}_T$, we have
     \bqn\label{17}
    \|U_{A_u}(t,s)-U_{A_{\bar{u}}}(t,s)\|_{\mathcal{L}(\Wqb^{2\alpha},
    \Wqb^{2\beta})}\,\le\,
    c(R)\, (t-s)^{\alpha-\beta}\,\|u-\bar{u}\|_{\mathcal{V}_T}
    \eqn
for $0\le s<t\le T$, $0\le \alpha, \beta\le 1$ with $\alpha\not=
0$, $\beta\not= 1$, $2\alpha, 2\beta \not= 1+1/p$.

Since $-\Delta_x$ subject to homogeneous Neumann conditions on the
boundary generates a contraction semigroup on $\Wqb^{2\sigma}$
according to
\cite[Corollary~V.2.1.4]{LQPP} it follows from \eqref{17} that
     \bqn\label{19}
    \|U_{A_u}(t,s)\|_{\mathcal{L}(\Wqb^{2\vr},\Wqb^{2\sigma})}\,\le\,
    c(R)\, (t-s)^{\vr-\sigma}+c_0
    \eqn
for $0\le s<t\le T$. Also note that
     \bqn\label{20}
    \|U_{A_u}(t,s)\|_{\mathcal{L}(L_r)}\,\le\,
    1\ ,\qquad r\in (1,\infty]\ ,\quad 0\le s\le t\le T\ .
    \eqn
Defining $\Phi$ by \eqref{16A} we now claim that
$\Phi:\mathcal{V}_T\rightarrow \mathcal{V}_T$ is a contraction
provided $T=T(R)\in (0,1)$ is chosen sufficiently small. To prove
this we fix $u\in\mathcal{V}_T$ and observe that $v_u\in
C^1([0,T],\Wqb^{2\sigma})$ is well-defined due to Lemma \ref{L1}.
Furthermore, Lemma \ref{L1} and \eqref{13} entail
    \bqn\label{21}
    \|\xi(v_u(t)) v_u(t)\|_{W_p^{2\sigma}}\,\le\, c(R)\ ,\quad 0\le
    t\le T\ .
    \eqn
We put $\lambda(a):={\bf 1}_{(a_0,\infty)}(a)\,e^{a/\tau}$ so that
$$
\Lambda_u(t,x) =\int_0^\infty \lambda(a)\, u(t,a,x)\,\rd a\ ,\quad
(t,x)\in [0,T]\times \Om\ .
$$
Then we deduce from \eqref{12}, \eqref{2000}, \eqref{16},
\eqref{18}, and \eqref{21} that, for $0\le s \le t \le T \le 1$,
    \bqnn
    \begin{split}
    \|\Lambda_{\Phi(u)}&(t)-\Lambda_{\Phi(u)}(s)\|_{W_p^{2\vartheta}}\\
    &\le\,
    \int_s^t \|U_{A_u}(t,a)\|_{\mathcal{L}(\Wqb^{2\vartheta})}\, \|\xi(v_u(a))
    v_u(a)\|_{W_p^{2\vartheta}}\,e^{(t-a)/\tau}\,\rd a\\
    &\quad+\int_0^s \left\vert e^{-\int_0^{t-a}\mu(r)\rd r}-
    e^{-\int_0^{s-a}\mu(r)\rd
    r}\right\vert\,
\|U_{A_u}(t,a)\|_{\mathcal{L}(\Wqb^{2\vartheta})}\,\|\xi(v_u(a))
    v_u(a)\|_{W_p^{2\vartheta}}\,e^{(t-a)/\tau}\,\rd a\\
    &\quad+\int_0^s
\|U_{A_u}(t,a)-U_{A_u}(s,a)\|_{\mathcal{L}(\Wqb^{2\sigma},\Wqb^{2\vartheta})}\,\|\xi(v_u(a))
v_u(a)\|_{W_p^{2\sigma}}\,e^{(t-a)/\tau}\,\rd a\\
     &\quad+\int_0^s
\|U_{A_u}(s,a)\|_{\mathcal{L}(\Wqb^{2\vartheta})}\,\|\xi(v_u(a))
    v_u(a)\|_{W_p^{2\vartheta}}\, \big\vert
\lambda(t-a)-\lambda(s-a)\big\vert\,\rd a\\
     &\quad+\|U_{A_u}(t,0)-U_{A_u}(s,0)
\|_{\mathcal{L}(\Wqb^{2\sigma},\Wqb^{2\vartheta})}\int_0^\infty
     \|u^0(a)\|_{W_p^{2\sigma}}\, e^{(a+t)/\tau}\,\rd a\\
       &\quad+\|U_{A_u}(s,0)\|_{\mathcal{L}(\Wqb^{2\vartheta})}\int_0^\infty
    \left\vert e^{-\int_a^{t+a}\mu(r)\rd r}- e^{-\int_a^{s+a}\mu(r)\rd
    r}\right\vert
     \|u^0(a)\|_{W_p^{2\vartheta}}\, e^{(a+t)/\tau}\,\rd a\\
     &\quad+\|U_{A_u}(s,0)\|_{\mathcal{L}(\Wqb^{2\vartheta})}\int_0^\infty
     \|u^0(a)\|_{W_p^{2\vartheta}}\,
\big\vert\lambda(a+t)-\lambda(a+s)\big\vert\,\rd a\\
     &\le\, c(R)\,(t-s)\,+\,c(R)\,
     (t-s)^{\sigma-\vartheta}\,+c(R)\, \int_0^s
     \big\vert\lambda(t-a)-\lambda(s-a)\big\vert\, \rd a\\
     &\quad+\, c(R)\int_0^\infty
     \|u^0(a)\|_{W_p^{2\vartheta}}\,
\big\vert\lambda(a+t)-\lambda(a+s)\big\vert\,\rd a\ .
     \end{split}
    \eqnn
Next note that
    \bqnn
    \begin{split}
    \int_0^\infty
     \|u^0&(a)\|_{W_p^{2\vartheta}}\,\big\vert\lambda(a+t) -
\lambda(a+s)\big\vert\,\rd a\\
     &\le\, \int_{(a_0-t)_+}^{(a_0-s)_+}
     \|u^0(a)\|_{W_p^{2\vartheta}}\,e^{(t+a)/\tau}\, \rd a
     \, + \int_{(a_0-s)_+}^\infty\|u^0(a)\|_{W_p^{2\vartheta}}\,
\left\vert e^{(a+t)/\tau}-e^{(a+s)/\tau}\right\vert\, \rd a\\
     &\le\, e^{(a_0+1)/\tau}\,
     \|u^0\|_{L_q((0,a_0),W_p^{2\vartheta})}\, \big\vert
     (a_0-s)_+ - (a_0-t)_+\big\vert^{1/q'}
     \, +\,
     \tau^{-2}\,e^{1/\tau}\, (t-s)
     \,\|u^0\|_{E_\vartheta}\\
     &\le\, c(R)\,\big( (t-s)^{1/q'}\, +\,(t-s) \big)
    \end{split}
    \eqnn
owing to \eqref{2000} while
    \bqnn
    \begin{split}
    \int_0^s\big\vert\lambda(t-a)-\lambda(s-a)\big\vert\, \rd a\, &\le\,
    \int_0^{(s-a_0)_+}\big(e^{(t-a)/\tau}-e^{(s-a)/\tau}\big)\,
    \rd a \,+\, \int_{(s-a_0)_+}^{(t-a_0)_+}
    e^{(t-a)/\tau}\, \rd a\\
    &\le\, \tau^{-2}\, e^{1/\tau}\, (t-s)\,+\,
    e^{1/\tau}\,\big( (t-a_0)_+-(s-a_0)_+\big)\\
    &\le\, \big( 1+\tau^{-2} \big)\, e^{1/\tau}\, (t-s)\ .
    \end{split}
    \eqnn
Therefore,
$$
\|\Lambda_{\Phi(u)}(t)-\Lambda_{\Phi(u)}(s)\|_{W_p^{2\vartheta}}\,\le\,
\vert t-s\vert^\kappa\ ,\quad 0\le t,s\le T\ ,
$$
due to the choice of $\kappa$ provided that $T=T(R)\in (0,1)$ is
chosen sufficiently small. Furthermore, using \eqref{16},
\eqref{19}, and \eqref{21} we obtain for $0\le t\le T$
    \bqnn
    \begin{split}
     \|\Phi(u)(t)\|_{E_\sigma}\,&\le\, \int_0^t
     \|U_{A_u}(t,t-a)\|_{\mathcal{L}(\Wqb^{2\sigma})}\,\|\xi(v_u(t-a))\,
     v_u(t-a)\|_{W_p^{2\sigma}}\, e^{a/\tau}\, \rd a\\
      &\quad +\int_t^\infty
     \|U_{A_u}(t,0)\|_{\mathcal{L}(\Wqb^{2\vr},\Wqb^{2\sigma})}\,
     \|u^0(a-t)\|_{W_p^{2\vr}}\, e^{a/\tau}\, \rd a\\
      &\le\, c(R)\,T\,+\, \big(c(R)\, t^{\vr-\sigma}\,+c_0\big)\,
     e^{1/\tau}\,\|u^0\|_{E_\vr}\\
     &\le \, 1\,+ \, R
    \end{split}
    \eqnn
provided that $T=T(R)\in (0,1)$ is chosen sufficiently small.
Since $\Phi(u)$ is obviously non-negative and $\Phi(u)\in
C([0,T],E_\sigma)$ holds by similar arguments as used to prove the
H\"older continuity of $\Lambda_{\Phi(u)}$, we conclude that
$\Phi$ maps $\mathcal{V}_T$ into itself. That it is a contraction
follows from the observation that if $u,\bar{u}\in \mathcal{V}_T$
and $0\le t\le T$, then
     \bqnn
    \begin{split}
     \|\Phi(u)&(t)-\Phi(\bar{u})(t)\|_{E_\sigma}\\
     & \le\, \int_0^t
     \|U_{A_u}(t,t-a)-U_{A_{\bar{u}}}(t,t-a)\|_{\mathcal{L}(\Wqb^{2\sigma})}\,\|\xi(v_u(t-a))\,
     v_u(t-a)\|_{W_p^{2\sigma}}\, e^{a/\tau}\, \rd a\\
     &\quad  +\int_0^t
     \|U_{A_{\bar{u}}}(t,t-a)\|_{\mathcal{L}(\Wqb^{2\sigma})}\,\|\xi(v_u(t-a))\,
     v_u(t-a) - \xi(v_{\bar{u}}(t-a))\,
     v_{\bar{u}}(t-a)\|_{W_p^{2\sigma}}\, e^{a/\tau}\, \rd a\\
     &\quad +\int_t^\infty
     \|U_{A_u}(t,0)-U_{A_{\bar{u}}}(t,0)\|_{\mathcal{L}(\Wqb^{2\vr},\Wqb^{2\sigma})}\,
     \|u^0(a-t)\|_{\Wqb^{2\vr}}\, e^{a/\tau}\, \rd a
    \end{split}
    \eqnn
and hence, using \eqref{13A}, \eqref{13}, \eqref{2000},
\eqref{16}, \eqref{17}, and \eqref{21},
    \bqnn
     \|\Phi(u)(t)-\Phi(\bar{u})(t)\|_{E_\sigma}\,\le\, c(R)\,
     \|u-\bar{u}\|_{\mathcal{V}_T} \int_0^t e^{a/\tau}\,\rd a
    \,+\, c(R)\, t^{\vr-\sigma}
     \,\|u-\bar{u}\|_{\mathcal{V}_T}
     \le\, \frac{1}{2}\, \|u-\bar{u}\|_{\mathcal{V}_T}
    \eqnn
provided that $T=T(R)\in (0,1)$ is chosen sufficiently small.
Therefore, by Banach's fixed point theorem there exists a unique
$u\in \mathcal{V}_T$ such that $\Phi(u)=u$. Note that \eqref{16}
and \eqref{21} imply $u=\Phi(u)\in L_\infty([0,T], E_1)$ since
$u^0\in E_1$, whence $v_u\in C([0,T],\Wqb^2)$ by Lemma \ref{L1}.
Due to $u(T)\in E_1$, $v_u(T)\in \Wqb^2$, and the fact that $T$
was chosen depending only on $R$ satisfying \eqref{2000}, we can
iterate this argument and extend $u$ and $v_u$ uniquely to
functions $u\in C(J,E_\sigma)\cap L_{\infty, loc}(J,E_1)$ and
$v\in C^1(J,\Wqb^{2\sigma})\cap C(J,\Wqb^2)$, where $t^+:=\sup
J=\infty$ if
    \bqn\label{30}
    \sup_{0<t<\min{\{t^+,T\}}}\big\{ \|u(t)\|_{E_\vr}\,+\,
    \|u(t)\|_{L_q((0,a_0),W_p^{2\omega})}\big\}\, <\, \infty \quad
    \text{for all}\quad T>0\ .
    \eqn
Clearly, this so extended function $u$ still satisfies
    \bqn\label{30A}
    u(t,a)\,=\,\left\{\begin{array}{ll}
    e^{-\int_0^a\mu(r)\rd r}\,
    U_{A}(t,t-a)\,\xi(v(t-a))\,v(t-a)\ ,& a<t\ ,\\
    e^{-\int_{a-t}^a\mu(r)\rd r}\,
    U_{A}(t,0)\,u^0(a-t)\ ,& a>t\
    \end{array}
    \right.
    \eqn
for $a>0$ and $0\le t<t^+$, where we simply write $A=A_u$ and
$v=v_u$. Next recall that $u^0\in E_1\cap C^1(\R^+,L_p)$ and so,
for $t\in (0,t^+)$ and $a>0$ with $a\not= t$,
    \bqnn
    \begin{split}
    \partial_t u(t,a)\,=\,& {\bf 1}_{[a<t]}(t,a)\, e^{-\int_0^a
    \mu(r)\rd r}\,\Big\{ -A(t)\, U_A(t,t-a)\,
    \xi(v(t-a))\,v(t-a)
    \\
    &\qquad \qquad\qquad\qquad\qquad  +\, U_A(t,t-a)\big(\partial_t\, +\,A(t-a)\big)\big(\xi(v(t-a))\,
    v(t-a)\big)\Big\}\\
    &- {\bf 1}_{[a>t]}(t,a)\,\left\{ \mu(a-t)\, u(t,a)\,+\, e^{-\int_{a-t}^a
    \mu(r)\rd r}\,A(t)\, U_A(t,0)\, u^0(a-t)\right.\\
    &\qquad \qquad\qquad\ \left. +\, e^{-\int_{a-t}^a
    \mu(r)\rd r}\,
    U_A(t,0)\,\partial_a u^0(a-t)
    \right\}
    \end{split}
    \eqnn
and
 \bqnn
    \begin{split}
    \partial_a u(t,a)\,=\,& -\mu(a)\, u(t,a)\\
    &-\, {\bf 1}_{[a<t]}(t,a)\, e^{-\int_0^a
    \mu(r)\rd r}\, U_A(t,t-a)\, \big(A(t-a)\,+\,\partial_t\big)\,\big(
    \xi(v(t-a))\,v(t-a)\big)\\
    &+ {\bf 1}_{[a>t]}(t,a)\,\left\{ e^{-\int_{a-t}^a
    \mu(r)\rd r}\, U_A(t,0)\,\partial_a
    u^0(a-t)\,+\, \mu(a-t)\, u(t,a) \right\}\ .
    \end{split}
    \eqnn
Thus $(v,u)$ is a solution to \eqref{1}-\eqref{5} with the
regularity properties as stated in the assertion of the theorem.

It remains to prove that $t^+=\infty$. We fix $T>0$ arbitrarily
and put $J_T:=J\cap [0,T]$. Defining $\M_1(t,x):=\int_0^\infty
e^{a/\tau}\,u(t,a,x)\,\rd a$ we observe that \eqref{2} and
\eqref{12} ensure
$$
\partial_t v(t)\,\le\, v(t)\,+\, \|\mu\|_\infty\, \M_1(t)\ ,\quad
t\in J\ ,
$$
whence
$$
\|v(t)\|_\infty\, \le\, c(T) \,\left(\int_0^t
\|\M_1(s)\|_\infty\,\rd s \,+\,1\right)\ ,\quad t\in J_T\ .
$$
But since
$$
\M_1(t)\,\le\, \|\xi\|_\infty\, e^{T/\tau}\int_0^t\|
v(t-a)\|_\infty\, \rd a\,+\, c(T)\,\|u^0\|_{E_1}\ ,\quad t\in J_T\
,
$$
by \eqref{20}, we conclude
    \bqn\label{31}
    \|v(t)\|_\infty\,+\, \|\M_1(t)\|_\infty\, \le\, c(T)\ , \quad
    t\in J_T\ .
    \eqn
Owing to $u^0\in C([0,a_0],W_p^{2\omega-1})$ and $v\in
C(J,\Wqb^2)$ it follows from
    \bqn\label{32}
    \|U_A(t,s)-U_A(t,r)\|_{\mathcal{L}(\Wqb^2,\Wqb^{2\omega-1})}\,\le
    c(t_0)\, (s-r)^{3/2-\omega}\ ,\quad 0\le r\le s\le t\le
    t_0<t^+\ ,
    \eqn
and \eqref{30A} that $u(\cdot,a_0)\in
C(J\setminus\{a_0\},W_p^{2\omega-1})$. Property \eqref{32} is
shown analogously to \cite[Equation (II.5.3.8)]{LQPP}. Provided
$a_0<t^+$, \eqref{18} warrants that $\lim_{t\nearrow a_0}
u(t,a_0)=u^0(0)$ in $W_p^{2\nu}$ with \mbox{$n/p<2\nu<2\omega-1$},
while \eqref{32} warrants that $\lim_{t\searrow a_0}
u(t,a_0)=\xi(v^0)v^0$ in $W_p^{2\omega-1}$. Thus, the imposed
compatibility condition on $u^0$ and $v^0$ entails
$u(\cdot,a_0)\in C(J,W_p^{2\nu})\hookrightarrow
C(J,C(\bar{\Om}))$. Recalling that $\Lambda_u$ is given by
\eqref{6} we set
$$
f(t,x)\,:=\, e^{a_0/\tau}\, u(t,a_0,x)\, +\,
\frac{1}{\tau}\,\Lambda_u (t,x) \, -\, \int_{a_0}^\infty
e^{a/\tau}\, \mu(a)\, u(t,a,x)\, \rd a\ ,\quad (t,x)\in
J_T\times\bar{\Om}\ ,
$$
and deduce $f\in C(J_T\times\bar{\Om})$ with
    \bqn\label{33}
    \vert f(t,x)\vert\, \le\, c(T)\ ,\quad (t,x)\in J_T\times\bar{\Om}
    \eqn
due to \eqref{31} and
$$
\|u(t,a_0)\|_\infty\, \le\, \left\{\begin{array}{ll}
\|\xi(v(t-a_0))\,v(t-a_0)\|_\infty\ ,&t>a_0\\
\|u^0(a_0-t)\|_\infty\ , &t<a_0 \end{array} \right\} \,\le \,
c(T)\ .
$$
We then observe that $\Lambda=\Lambda_u$ solves the quasilinear
parabolic problem
$$
    \partial_t\Lambda\,-\, \divv_x\big(
    D(\Lambda)\nabla_x\Lambda\big)\,=\, f(t,x)\ , \quad (t,x)\in
    J_T\times\bar{\Om}
$$
subject to $\partial_\nu \Lambda =0$ and $ \Lambda(0)\in\Wqb^2$.
We refer to \eqref{10}, \eqref{31}, and \eqref{33} when using
\cite[Lemma~5.1(ii)]{AmannIII} to obtain that $\Lambda\in
BUC^\delta(J_T,C^\delta(\bar{\Om}))$ for some $\delta>0$, and
hence $\Lambda\in BUC^\varepsilon (J_T,C^1(\bar{\Om}))$ for some
$\varepsilon>0$ by \cite[Lemma~4.2, Remark~4.3]{AmannIII}, where
$BUC^\varepsilon$ stands for `bounded and uniformly $\varepsilon
$-H\"older continuous'. But then $A=A_u$ is uniformly H\"older
continuous, that is,
$$
\|A_u(t)-A_u(s)\|_{\mathcal{L}(\Wqb^2,L_p)}\,\le\, c(T)\, \vert
t-s\vert^\varepsilon\ ,\quad t,s\in J_T\ ,
$$
so that \cite[Lemma~II.5.1.3]{LQPP} implies
     \bqn\label{400}
    \|U_{A_u}(t,s)\|_{\mathcal{L}(\Wqb^{2\vr})}\,+\,
    (t-s)^{\vr}\,\|U_{A_u}(t,s)\|_{\mathcal{L}(L_p,\Wqb^{2\vr})}\,\le\,
    c(T)
    \eqn
for $t,s\in J_T$ with $s<t$. Note that $c(T)$ depends here on $T$
only (but not on some norm of $u$). Combining \eqref{30A},
\eqref{31}, and \eqref{400} we have
    \bqnn
    \begin{split}
    \|u(t)\|_{L_q((0,a_0),W_p^{2\omega})}^q\,&\le\,
    \int_0^{\min{\{t,a_0}\}}\|U_A(t,t-a)\|_{\mathcal{L}(L_p,\Wqb^{2\omega})}^q\,
    \,\|\xi(v(t-a))\,v(t-a)\|_{L_p}^q\,\rd a\\
    &\quad +\int_{\min{\{t,a_0}\}}^{a_0}
    \|U_A(t,0)\|_{\mathcal{L}(\Wqb^{2\omega})}^q\,
    \|u^0(a-t)\|_{W_p^{2\omega}}^q\, \rd a\\
    &\le\, c(T)\int_0^{\min{\{t,a_0}\}} a^{-q\omega}\, \rd a\,+\, c(T)
    \|u^0\|_{L_q((0,a_0),W_p^{2\omega})}\\
    &\le \, c(T)
    \end{split}
    \eqnn
for $t\in J_T$ thanks to $q\omega<1$. Finally, from \eqref{30A},
\eqref{31}, and \eqref{400} it follows analogously that $u\in
L_\infty (J_T, E_\vr)$. From this and \eqref{30} we deduce
$t^+=\infty$. This proves the theorem.
\end{proof}

\section{The Degenerate Case}\label{sect3}

We now turn to the ``degenerate'' case where $D$ is allowed to
vanish but only for $\Lambda=0$. More precisely, we assume that
$D\in C^{2-}(\R)$ is such that $D(0)=0$, $D(z)>0$ if $z>0$, and
\bqn \label{z0} I_D:= \int_0^1 \frac{z\ D'(z)^2}{D(z)}\, \rd z <
\infty \;\;\mbox{ and }\;\; \lim_{z\to 0}
\frac{z\,|D'(z)|}{D(z)^{1/2}} = 0\ . \eqn The function \bqn
\label{z00} \Phi_D(z):= z\ \int_1^z \frac{D'(y)^2}{D(y)}\, \rd y -
\int_0^z \frac{y\ D'(y)^2}{D(y)}\, \rd y\ , \quad z\in [0,\infty)\
, \eqn is then a well-defined smooth convex function satisfying
$\Phi_D(z)\ge \Phi_D(1)=-I_D$ for $z\in [0,\infty)$. We also put
$$
\widehat{D}(z) \,:=\, \int_0^z D(y)\, \rd y \;\;\mbox{ and }\;\;
\widehat{D}_1(z) \,:=\, \int_0^z \widehat{D}(y)\, \rd y\ , \quad
z\in [0,\infty)\ .
$$
We note that both $D(z)=z^{m-1}$, $m>1$ and $D(z)=e^{-1/z}$ fulfil
\eqref{z0}.

As for the differentiation and dedifferentiation rates $\xi$ and
$\mu$ we assume that $\xi$ fulfils \eqref{11} while $\mu$
satisfies \bqn \label{z000} \mu\in W_\infty^1(\R) \;\;\mbox{ is
non-negative and }\;\; \mu(a) = 0 \;\;\mbox{ for }\;\; a<a_0\ ,
\eqn the latter assumption being stronger than \eqref{12}.

Finally, the initial data are required to satisfy the following
properties: \bqn \label{zhid1} 0\le u^0\in
L_1\big((0,\infty)\times\Om, e^{a/\tau}\,\rd a\rd x \big)\cap
L_\infty((0,\infty)\times\Om) \;\;\mbox{ and  }\;\; 0\le v^0 \in
L_\infty(\Om) \eqn and \bqn \label{zhid3} \Lambda^0\in
L_\infty(\Om) \;\;\mbox{ and }\;\; \widehat{D}\left( \Lambda^0
\right)\in W_2^1(\Om)\ , \eqn where \bqn \label{zhid2}
\Lambda^0(x)\,:=\,\int_{a_0}^\infty e^{a/\tau}\, u^0(a,x)\, \rd a\
, \quad x\in\Om\ . \eqn

\medskip

\begin{thm}\label{wsdgn} Let $T>0$ and put $U\,:=\,(0,T)\times
(0,\infty)\times\Om$. There are two non-negative functions $u\in
L_\infty(U)$ and $v\in L_\infty((0,T)\times\Om) \cap
C([0,T],L_2(\Om))$ satisfying
$$
\partial_t v(t,x) \,=\, \frac{1}{\tau}\,\big(1\,-\,\xi(v(t,x))\big)\,
v(t,x)\,+\int_0^\infty e^{a/\tau}\,\mu(a)\, u(t,a,x)\, \rd a
\;\;\mbox{ a.e. in }\;\; (0,T)\times\Om\ ,
$$
and \bean 0 & = & \int_0^T \int_\Om \int_0^\infty u(t,a,x)\, \big(
\partial_t\psi\,+\,\partial_a \psi \big)(t,a,x)\,\rd a\rd
x\rd t + \int_\Om \int_0^\infty u^0(a,x)\,\psi(0,a,x)\,\rd a\rd x \\
& + & \frac{1}{\tau}\, \int_0^T \int_\Om \xi(v(t,x))\, v(t,x)\,
\psi(t,0,x)\,\rd x\rd t - \int_0^T \int_\Om \int_0^\infty
\mu(a)\,u(t,a,x)\,\psi(t,a,x)\,\rd a\rd x\rd t \\
& - &  \int_0^T \int_\Om \int_0^\infty
\mathbf{J}(t,a,x)\, \nabla_x \psi(t,a,x)\,\rd a\rd x\rd t \eean
for $\psi\in C_c^1([0,T)\times [0,\infty)\times\bar{\Omega})$ with
$\partial_\nu \psi(t,a,x)=0$ for $(t,a,x)\in (0,T)\times
(0,\infty)\times\partial\Om$, where the functions $\Lambda$ and
$\mathbf{J}$ are given by
\begin{eqnarray}
\label{fantasio}
\Lambda(t,x) & := & \int_{a_0}^\infty e^{a/\tau}\,u(t,a,x)\,\rd a
\;\;\mbox{ for a.e. }\;\; (t,x)\in (0,T)\times\Om\ , \\ 
\label{spirou} \mathbf{J} & := & \nabla_x\left( u\,D(\Lambda)
\right)\,-\, u\,\nabla_x D(\Lambda) \;\;{ in }\;\;
\mathcal{D}'(U,\R^n)\ ,
\end{eqnarray}
and satisfy $\Lambda\in C([0,T],L_2(\Om))$, $D(\Lambda)\in
L_2((0,T),W_2^1(\Om))$, and $\mathbf{J}\in L_2(U,\R^n)$.
\end{thm}

The proof of Theorem~\ref{wsdgn} is performed by a compactness
argument, approximating the diffusivity $D$ by non-degenerate
diffusivities $(D_\alpha)_{\alpha>0}$ for which we can apply
Theorem~\ref{T2} and obtain a sequence of solutions
$(v_\alpha,u_\alpha)_{\alpha>0}$. The next step is to pass to the
limit as $\alpha\to 0$ and we now point out the difficulties to be
overcome: first, as $D(\Lambda)$ vanishes when $\Lambda=0$, the
equation \eqref{1} is no longer uniformly parabolic with respect
to the space variable and $\nabla_x u$ is unlikely to be a
function. Furthermore, as $a_0>0$, we may have $\Lambda(t,x)=0$
but $u(t,a,x)\not= 0$ for $a\in (0,a_0)$ and \eqref{1} gives no
information on $u$ in that case. We therefore cannot expect to
have strong convergence on the sequence $(u_\alpha)$. There are
however nonlinear terms in \eqref{1} and \eqref{2} for which
strong convergence is necessary to identify the limit. In
particular, the strong compactness of $(v_\alpha)_\alpha$ is
needed to pass to the limit in the term $\xi(v_\alpha)\,v_\alpha$.
As $v_\alpha$ solves an ordinary differential equation, such a
compactness can only be obtained as a consequence of that of
$(t,x)\longmapsto \int_0^\infty
e^{a/\tau}\,\mu(a)\,u_\alpha(t,a,x)\,\rd a$. One step in the proof
is thus to show that certain integrals of $u_\alpha$ with respect
to age enjoy some compactness properties with respect to the time
and space variables. The strong compactness of
$\Lambda_\alpha(t,x)=\int_{a_0}^\infty
e^{a/\tau}\,u_\alpha(t,a,x)\,\rd a$ will also follow from this
step. Next, in order to identify $\mathbf{J}$, strong compactness
is needed on $(\nabla_x D(\Lambda_\alpha))_\alpha$ to pass to the
limit in the term $u_\alpha \nabla_x D(\Lambda_\alpha)$. This is
proved by a suitable adaptation of an argument from \cite{AMST99}.

\bigskip

We now begin the proof of Theorem~\ref{wsdgn}. We fix $T>0$ and
consider a sequence $(D_\alpha)_{\alpha\in (0,1)}$ of functions in
$C^{2-}(\R)$ with the following properties: for every $\alpha\in
(0,1)$, there is $d_\alpha>0$ such that $D_\alpha(z)\ge d_\alpha$
for all $z\in\R$, and \bqn \label{zz0} D_\alpha(z) = D(z)
\;\;\mbox{ for }\;\ z \in \big[ \alpha\,e^{-T\,\|\mu\|_\infty} ,
\infty \big)\ . \eqn Next, let $(v_\alpha^0,u_\alpha^0)_{\alpha\in
(0,1)}$ be a sequence of non-negative initial data fulfilling all
the requirements of Theorem~\ref{T2} together with the following
properties: \bqn \label{zz1} \lim_{\alpha\to 0} \left\{ \int_\Om
\int_0^\infty e^{a/\tau}\,\left| u_\alpha^0(a,x) - u^0(a,x)
\right|\,\rd a\rd x + \big\| v_\alpha^0 - v^0 \big\|_1 \right\} =
0\ , \eqn and there is $c_0>0$ such that \bqn \label{zz2} \int_\Om
\int_0^\infty e^{a/\tau}\,u_\alpha^0(a,x)\,\rd a\rd x + \left\|
u_\alpha^0 \right\|_\infty + \left\| v_\alpha^0 \right\|_\infty +
\left\|\Lambda_\alpha^0\right\|_\infty + \big\| \widehat{D}\left(
\Lambda_\alpha^0 \right)\big\|_{W_2^1(\Om)} \le c_0 \eqn with
    \bqn\label{AA}
\Lambda_\alpha^0(x)\,:=\,\int_{a_0}^\infty e^{a/\tau}\,
u_\alpha^0(a,x)\, \rd a \ge \alpha
    \eqn
for all $x\in\Om$ and $\alpha\in (0,1)$.

We denote by $(v_\alpha,u_\alpha)$ the solution to
 \begin{align}
\partial_t u_\alpha+\partial_a u_\alpha&\,=\, \divv_x
\big( D_\alpha(\Lambda_\alpha)\,\nabla_x u_\alpha \big)\,
-\,\mu(a)\, u_\alpha\ , \qquad  (t,a,x)\in
(0,\infty)\times(0,\infty)\times\Om\
,\label{zz3}\\
\partial_t v_\alpha&\,=\, \frac{1}{\tau}\,\big(1\,-\,\xi(v_\alpha)\big)\,
v_\alpha\,+\int_0^\infty e^{a/\tau}\,\mu(a)\, u_\alpha(t,a,x)\,
\rd a\ ,\ (t,x)\in (0,\infty)\times\Om\ ,\label{zz4}
\end{align}
where \bqn\label{zz5}
    \Lambda_\alpha(t,x)\,:=\int_{a_0}^\infty e^{a/\tau}\,
u_\alpha(t,a,x)\, \rd a\ ,\quad (t,x)\in (0,\infty)\times\Om\ ,
    \eqn
subject to the boundary conditions
\begin{align}
u_\alpha(t,0,x)&\,=\,
\frac{1}{\tau}\,\xi(v_\alpha(t,x))\,v_\alpha(t,x)\ ,\quad
&(t,x)\in(0,\infty)\times\Om\ ,\label{zz6}\\
D_\alpha(\Lambda_\alpha)\,\partial_\nu\, u_\alpha\,&\,=\,0\ ,
&(t,x)\in (0,\infty)\times \partial\Om\ ,\label{zz7}
\end{align}
and the initial conditions
    \bqn\label{zz8}
 u_\alpha(0,a,x)=u_\alpha^0(a,x)\ ,\quad
v_\alpha(0,x)=v_\alpha^0(x)\ ,\qquad (a,x)\in (0,\infty)\times\Om\
.
    \eqn
We note that, thanks to \eqref{zz3}, $\Lambda_\alpha$ solves \bqn
\label{zz9}
\partial_t \Lambda_\alpha \,=\, \divv_x
\big( D_\alpha(\Lambda_\alpha)\,\nabla_x \Lambda_\alpha \big)\, +
g_\alpha^1 - g_\alpha^2 \;\;\mbox{ in }\;\; (0,T)\times\Om \eqn
with homogeneous Neumann boundary conditions and
$$
g_\alpha^1(t,x) \,:=\, e^{a_0/\tau}\, u_\alpha(t,a_0,x) +
\frac{\Lambda_\alpha(t,x)}{\tau}\ge 0 \;\;\mbox{ and }\;\;
g_\alpha^2(t,x) \,:=\, \int_{a_0}^\infty \mu(a)\,e^{a/\tau}\,
u_\alpha(t,a,x)\ \rd a \ge 0\ .
$$

For further use, we introduce the following functions:
$$
\M_{1,\alpha}(t,x) := \int_0^\infty e^{a/\tau}\, u_\alpha(t,a,x)\,
\rd a \;\;\mbox{ and }\;\; \M_{2,\alpha}(t,x) := \int_0^\infty
e^{a/\tau}\,\mu(a)\, u_\alpha(t,a,x)\, \rd a
$$
for $(t,x)\in (0,T)\times\Om$ and $\alpha\in (0,1)$. Using again
\eqref{zz3} together with \eqref{zz5}, we realize that
$\M_{1,\alpha}$ and $\M_{2,\alpha}$ are solutions to \bqn
\label{zz10}
\partial_t \M_{1,\alpha} \,=\, \divv_x
\big( D_\alpha(\Lambda_\alpha)\,\nabla_x \M_{1,\alpha} \big)\, +
\frac{1}{\tau}\,\xi(v_\alpha)\,v_\alpha +
\frac{\M_{1,\alpha}}{\tau}\,-\,\int_0^\infty \mu(a)\,e^{a/\tau}\,
u_\alpha(.,a,.)\ \rd a \eqn and \bqn \label{zz11}
\partial_t \M_{2,\alpha} \,=\, \divv_x
\big( D_\alpha(\Lambda_\alpha)\,\nabla_x \M_{2,\alpha} \big)\, +
\, \frac{\M_{2,\alpha}}{\tau}\, \,+\, \int_0^\infty \left(
\mu'(a)\,-\,\mu(a)^2 \right)\, e^{a/\tau}\, u_\alpha(.,a,.)\ \rd
a\ \eqn in $(0,T)\times\Om$, respectively, with homogeneous
Neumann boundary conditions.

\medskip

In the following, we denote by $c$ and $c_i$, $i\ge 1$, positive
constants depending on $D$, $\mu$, $\xi$, $a_0$, $\tau$, and $c_0$
in \eqref{zz2}, but not on $\alpha$. The dependence upon
additional variables (such as $T$) will be indicated explicitly.
As in the non-degenerate case, we establish $L_\infty$-bounds for
$\M_{1,\alpha}$, $v_\alpha$, and $u_\alpha$.

\begin{lem}\label{lez1}
For $\alpha\in (0,1)$ and $t\in [0,T]$, we have
\begin{eqnarray}
\label{zz12} \| \M_{1,\alpha}(t)\|_\infty + \|
v_\alpha(t)\|_\infty + \|
u_\alpha(t)\|_\infty & \le & c_1(T)\ ,\\
\label{zz12a} \int_0^T \int_\Om \int_0^\infty
D_\alpha(\Lambda_\alpha)(t,x)\, |\nabla_x u_\alpha(t,a,x)|^2\,\rd
a\rd x\rd t & \le & c_1(T)\ .
\end{eqnarray}
\end{lem}

\begin{proof}
On the one hand, since $\mu$ belongs to $L_\infty(0,\infty)$, we
have
$$
\partial_t v_\alpha \le \frac{1}{\tau}\, v_\alpha + \|\mu\|_\infty\,
\M_{1,\alpha}
$$
by (\ref{zz4}), from which we deduce that, for $t\in [0,T]$,
$$
v_\alpha(t,x) \le v_\alpha^0(x)\ e^{t/\tau} + \|\mu\|_\infty\,
\int_0^t \M_{1,\alpha}(s,x)\,e^{(t-s)/\tau}\, \rd s\ .
$$
Using \eqref{zz2} gives \bqn \label{zz13} \|v_\alpha(t)\|_\infty
\le c(T)\, \left( 1 + \int_0^t \| \M_{1,\alpha}(s)\|_\infty\, \rd
s \right)\ . \eqn On the other hand, it follows from \eqref{zz10}
and the boundedness of $\xi$ that
$$
\partial_t \M_{1,\alpha} \,\le\, \divv_x
\big( D_\alpha(\Lambda_\alpha)\,\nabla_x \M_{1,\alpha} \big)\, +
\frac{\|v_\alpha\|_\infty}{\tau} + \frac{\M_{1,\alpha}}{\tau}\ .
$$
The comparison principle then ensures that
$$
\M_{1,\alpha}(t,x) \le \|\M_{1,\alpha}(0)\|_\infty\ e^{t/\tau} +
\frac{1}{\tau}\, \int_0^t \|v_\alpha(s)\|_\infty\,e^{(t-s)/\tau}\,
\rd s \le c(T)\, \left( 1 + \int_0^t \|v_\alpha(s)\|_\infty\, \rd
s \right)
$$
for $t\in [0,T]$. We now combine this estimate with \eqref{zz13}
and end up with
$$
\| \M_{1,\alpha}(t)\|_\infty \le c(T)\, \left( 1 + \int_0^t
\int_0^s \| \M_{1,\alpha}(\sigma)\|_\infty\, \rd \sigma \rd s
\right) \le  c(T)\, \left( 1 + \int_0^t \|
\M_{1,\alpha}(s)\|_\infty\, \rd s \right)\
$$
for $t\in [0,T]$. The Gronwall lemma then gives the claimed bound
on $\|\M_{1,\alpha}\|_\infty$, which in turn gives that for
$\|v_\alpha\|_\infty$ by \eqref{zz13}.

Finally, by \eqref{zz3}, \eqref{zz6}, and \eqref{zz7}, $u_\alpha$
satisfies
$$
\partial_t u_\alpha + \partial_a u_\alpha \le \divv_x
\big( D_\alpha(\Lambda_\alpha)\,\nabla_x u_{\alpha} \big)
$$
with $u_\alpha(t,0,x)=\xi(v_\alpha(t,x))\, v_\alpha(t,x)\,/\,\tau$
and subject to homogeneous Neumann boundary conditions for
\mbox{$t\in [0,T]$}. On the one hand, the comparison principle
readily implies that
$$
u_\alpha(t,a,x) \le \|u_\alpha^0\|_\infty + \frac{1}{\tau}\,
\sup_{s\in [0,T]}{\|v_\alpha(s)\|_\infty}\ , \quad (t,a,x)\in
(0,T)\times (0,\infty)\times\Om\ ,
$$
which, together with \eqref{zz2} and the already established bound
on $\|v_\alpha\|_\infty$, allows us to complete the proof of
\eqref{zz12}. On the other hand, since $u_\alpha$ is non-negative,
we also have \bean & & \int_\Om \int_0^\infty u_\alpha(T)^2\,\rd
a\rd x + 2\, \int_0^T \int_\Om \int_0^\infty
D_\alpha(\Lambda_\alpha(t))\,|\nabla_x u_\alpha(t)|^2\,\rd a\rd
x\rd t \\
& \le & \int_\Om \int_0^\infty \left( u_\alpha^0 \right)^2\,\rd
a\rd x + \int_0^T \int_\Om \int_0^\infty u_\alpha(t)^2\,\rd a\rd x
\rd t + \int_0^T \int_\Om \left( \xi(v_\alpha(t))\,v_\alpha(t)
\right)^2\, \rd
x\rd t \\
& \le & \|u_\alpha^0\|_\infty\,\|u_\alpha^0\|_1 + \int_0^T
\|u_\alpha(t)\|_\infty\,\|u_\alpha(t)\|_1\,\rd t + \int_0^T
\|v_\alpha(t)\|_\infty^2\,\rd t \\
& \le & c(T)\ , \eean the last inequality being a consequence of
\eqref{zz2}, \eqref{zz12}, and the obvious inequality
$\|u_\alpha(t)\|_1\le\|\M_{1,\alpha}(t)\|_1$ for $t\in [0,T]$.
\end{proof}

\medskip

We next derive some estimates for $\Lambda_\alpha$.

\begin{lem}\label{lez2}
For $\alpha\in (0,1)$, $t\in [0,T]$, and $x\in\Om$, we have
\begin{eqnarray}
\label{zz14} \|\Lambda_\alpha(t)\|_\infty + \int_0^T \int_\Om
\big\vert\nabla_x D(\Lambda_\alpha)\big\vert^2\, \rd x \rd s +
\int_0^T \int_\Om
\frac{D(\Lambda_\alpha)}{\Lambda_\alpha}\,\big|\nabla_x
\Lambda_\alpha\big|^2\,
\rd x \rd s & \le & c_2(T)\ , \\
\label{zz15} \big\| \widehat{D}(\Lambda_\alpha)(t)
\big\|_{W_2^1(\Om)} + \int_0^T \big\| \partial_t
\widehat{D}(\Lambda_\alpha)(s) \big\|_2^2\,\rd s & \le & c_2(T)\ ,
\end{eqnarray}
and $\Lambda_\alpha(t,x)\ge \alpha\,e^{-T\,\|\mu\|_{\infty}}$.
\end{lem}

A straightforward consequence of \eqref{zz0} and the last
assertion of Lemma~\ref{lez2} is that
$$
D_\alpha(\Lambda_\alpha)(t,x) = D(\Lambda_\alpha)(t,x) \;\;\mbox{
for }\;\; (t,x)\in (0,T)\times\Om\ .
$$

\begin{proof}
Clearly $\Lambda_\alpha\le \M_{1,\alpha}$ and the $L_\infty$-bound
for $\Lambda_\alpha$ is a straightforward consequence of
Lemma~\ref{lez1}. It next follows from \eqref{zz9} that
$\partial_t\Lambda_\alpha \,\ge\, \divv_x \big(
D_\alpha(\Lambda_\alpha)\,\nabla_x \Lambda_\alpha \big)\, - \,
\|\mu\|_\infty\,\Lambda_\alpha$ in $(0,T)\times\Om$ with
homogeneous Neumann boundary conditions. As $t\mapsto
\alpha\,e^{-t\,\|\mu\|_{\infty}}$ is a subsolution to the previous
equation, the lower bound $\Lambda_\alpha\ge
\alpha\,e^{-T\,\|\mu\|_{\infty}}$ in $(0,T)\times\Om$ readily
follows from \eqref{AA} by the comparison principle.

We next multiply \eqref{zz9} by $\Phi_D'(\Lambda_\alpha)$ with
$\Phi_D$ being defined in \eqref{z00} and integrate over
$(0,T)\times\Omega$ to obtain \bean & & \int_\Om \left(
\Phi_D(\Lambda_\alpha(t)) - \Phi_D(\Lambda_\alpha^0) \right)\, \rd
x + \int_0^T \int_\Om \Phi_D''(\Lambda_\alpha)\,
D(\Lambda_\alpha)\,\left|\nabla_x\Lambda_\alpha\right|^2\, \rd
x \rd t \\
& = & \int_0^T \int_\Om g_\alpha^1\, \Phi_D'(\Lambda_\alpha)\, \rd
x\rd t \,-\, \int_0^T \int_\Om g_\alpha^2\,
\Phi_D'(\Lambda_\alpha)\, \rd x\rd t \ . \eean On the one hand,
since $\Phi_D'$ is non-positive in $(0,1)$ and $g_\alpha^1\ge 0$, we
infer from \eqref{zz12} and the $L_\infty$-estimate on
$\Lambda_\alpha$ that \bean
 \int_0^T \int_\Om g_\alpha^1\, \Phi_D'(\Lambda_\alpha)\,
\rd x\rd t & \le & \int_0^T \int_\Om g_\alpha^1\,
\mathbf{1}_{[1,\infty)}(\Lambda_\alpha)\,
\Phi_D'(\Lambda_\alpha)\, \rd x\rd t \\
& \le & \int_0^T \int_\Om \left( e^{a_0/\tau}\,\|u_\alpha\|_\infty
+ \frac{\|\Lambda_\alpha\|_\infty}{\tau} \right)\, \Phi_D'\big( 1
+
\|\Lambda_\alpha\|_\infty \big)\,\rd x\rd t \\
& \le & c(T)\,. \eean On the other hand, as $\Phi_D'\ge 0$ on
$(1,\infty)$, $\Phi_D'\le 0$ on $(0,1)$, and $r
\left|\Phi_D'(r)\right| \le I_D$ for $r\in [0,1]$, we have \bean
-\,\int_0^T \int_\Om g_\alpha^2\, \Phi_D'(\Lambda_\alpha)\, \rd
x\rd t & \le & \int_0^T \int_\Om \int_{a_0}^\infty \mu(a)\,
e^{a/\tau}\, u_\alpha(t,a,x)\ \rd a\,
\mathbf{1}_{(0,1)}(\Lambda_\alpha)\,\left|\Phi_D'(\Lambda_\alpha)\right|\,
\rd x\rd t \\
& \le & \|\mu\|_\infty\, \int_0^T \int_\Om \Lambda_\alpha\,
\mathbf{1}_{[0,1)}(\Lambda_\alpha)\,\left|\Phi_D'(\Lambda_\alpha)\right|\,
\rd x\rd t \\
& \le & \|\mu\|_\infty\, I_D\, T \, |\Omega|\ . \eean Recalling
that $\Phi_D(r)\ge - I_D$ for $r\ge 0$ and $D\,\Phi_D''\, = \,
(D')^2$, we conclude that
$$
\int_0^T \int_\Om \left| \nabla_x D(\Lambda_\alpha) \right|^2\,
\rd x\rd t \le c(T) + \int_\Om \Phi_D(\Lambda_\alpha(0))\,\rd x
\le c(T) + \max{\left\{ I_D , \Phi_D\left( c_0 \right)
\right\}}\,.
$$

Similarly, we multiply \eqref{zz9} by $\log{\Lambda_\alpha}$ and
integrate over $(0,T)\times\Om$: using the non-negativity of
$g_\alpha^1$ and $g_\alpha^2$, \eqref{zz2}, and the
$L_\infty$-bound on $\Lambda_\alpha$ we obtain \bean & & \int_0^T
\int_\Om \frac{D(\Lambda_\alpha)}{\Lambda_\alpha}\,\left| \nabla_x
\Lambda_\alpha \right|^2\, \rd
x\rd t \\
& \le & \int_\Om \Lambda_\alpha^0\,\left( \log{\Lambda_\alpha^0}\,-\,1 \right)\,\rd x - \int_\Om \Lambda_\alpha(T)\,\big( \log{\Lambda_\alpha(T)}\,-\,1 \big)\,\rd x \\
& + & \int_0^T \int_\Om g_\alpha^1\,\log{\Lambda_\alpha}\,\mathbf{1}_{[1,\infty)}(\Lambda_\alpha)\,\rd x\rd t - \int_0^T \int_\Om g_\alpha^2\,\log{\Lambda_\alpha}\,\mathbf{1}_{(0,1)}(\Lambda_\alpha)\,\rd x\rd t \\
& \le & c(T) + T\,|\Om|\,\|g_\alpha^1\|_\infty\,\log{\big(1+\|\Lambda_\alpha\|_\infty\big)} + \|\mu\|_\infty\, \int_0^T \int_\Om \Lambda_\alpha\,|\log{\Lambda_\alpha}|\,\mathbf{1}_{(0,1)}(\Lambda_\alpha)\,\rd x\rd t \\
& \le & c(T) \eean as $\|g_\alpha^1\|_\infty$ is bounded uniformly
with respect to $\alpha\in (0,1)$ by \eqref{zz12} and the
$L_\infty$-bound on $\Lambda_\alpha$.

\medskip

We next multiply \eqref{zz9} by $2\,\partial_t
\widehat{D}(\Lambda_\alpha)$ and integrate over
$(0,t)\times\Omega$, $t\in [0,T]$: using \eqref{zz2},
\eqref{zz12}, and \eqref{zz14} we obtain \bean & & 2\,\int_0^t
\int_\Om D(\Lambda_\alpha)\,|\partial_t\Lambda_\alpha|^2\,\rd x\rd
s \,+\,
\big\| \nabla_x\widehat{D}(\Lambda_\alpha)(t) \big\|_2^2\\
& \le & \big\| \nabla_x\widehat{D}(\Lambda_\alpha^0) \big\|_2^2
\,+\, 2\,\int_0^t \int_\Om
D(\Lambda_\alpha)\,|\partial_t\Lambda_\alpha|\, \left(
e^{a_0/\tau}\, u_\alpha(s,a_0,x) + \left( \frac{1}{\tau} +
\|\mu\|_\infty \right)\,
\Lambda_\alpha \right)\,\rd x\rd s \\
& \le & c(T) \left( 1 \,+ \,\int_0^t \int_\Om
D(\Lambda_\alpha)\,|\partial_t\Lambda_\alpha|\,\rd x\rd s \right) \\
& \le & \int_0^t \int_\Om
D(\Lambda_\alpha)\,|\partial_t\Lambda_\alpha|^2\,\rd x\rd s +
c(T)\ . \eean Therefore
$$
\int_0^t \int_\Om
D(\Lambda_\alpha)\,|\partial_t\Lambda_\alpha|^2\,\rd x\rd s \,+\,
\big\| \nabla_x\widehat{D}(\Lambda_\alpha)(t) \big\|_2^2 \le
c(T)\, ,
$$
from which the claim \eqref{zz15} follows as $\big|
\partial_t\widehat{D}(\Lambda_\alpha) \big|\le
c(T)\,\sqrt{D(\Lambda_\alpha)}\,\left| \partial_t\Lambda_\alpha
\right|$ by \eqref{zz14}.
\end{proof}

\medskip

At this point we have gathered the information required to show
the strong compactness of $(\Lambda_\alpha)_\alpha$. This is,
however, not sufficient to pass to the limit as $\alpha\to 0$ as
there is a nonlinear dependence on $v_\alpha$ in \eqref{zz4}. We
now aim at proving the strong compactness of $(v_\alpha)$: this
will be achieved by the strong compactness of
$(\M_{2,\alpha})_\alpha$ which we show now.

\begin{lem}
\label{lez3} For $\alpha\in (0,1)$, $t\in [0,T]$, and $\delta\in
(0,1)$, we have \bqn \label{zz16} \int_0^T \left(
\big\|\nabla_x\left( \M_{2,\alpha} - \delta \right)_+^2 \big\|_2^2
+ \big\| \partial_t\left( \M_{2,\alpha} - \delta \right)_+^2
\big\|_{W_{n+1}^1(\Om)'} \right)\, \rd t \le c_3(T,\delta)\ . \eqn
\end{lem}

\begin{proof}
We multiply \eqref{zz11} by $\left( \M_{2,\alpha} - \delta
\right)_+$ and integrate over $(0,T)\times\Om$ to obtain \bean & &
\frac{1}{2}\,\big\| \left( \M_{2,\alpha}(T) - \delta \right)_+
\big\|_2^2 + \int_0^T \int_\Om D(\Lambda_\alpha)\, \big|
\nabla_x\left( \M_{2,\alpha} - \delta \right)_+ \big|^2\, \rd x
\rd t\\
& \le &  \frac{1}{2}\,\big\| \left( \M_{2,\alpha}(0) - \delta
\right)_+ \big\|_2^2 + T\,|\Om|\,\left(
\frac{\|\M_{2,\alpha}\|_\infty}{\tau} +
\|\mu'\|_\infty\,\|\M_{1,\alpha}\|_\infty \right)\, \big\| \left(
\M_{2,\alpha} - \delta \right)_+ \big\|_\infty\ . \eean As $\left(
\M_{2,\alpha} - \delta \right)_+\,\le\,\M_{2,\alpha}\,\le\,
\|\mu\|_\infty\, \M_{1,\alpha}$ and $\mu\in W_\infty^1(0,\infty)$,
we infer from Lemma~\ref{lez1} that \bqn \label{zz17} \int_0^T
\int_\Om D(\Lambda_\alpha)\, \big| \nabla_x\left( \M_{2,\alpha} -
\delta \right)_+ \big|^2\, \rd x \rd t \le c(T)\ . \eqn

Now, on the one hand, since the support of $\mu$ is included in
$[a_0,\infty)$, we have $\M_{2,\alpha}\,\le\, \|\mu\|_\infty\,
\Lambda_\alpha$ and
$$
\big\{ (t,x)\in (0,T)\times\Om\;:\;\M_{2,\alpha}(t,x)\ge\delta
\big\} \subset \big\{ (t,x)\in
(0,T)\times\Om\;:\;\Lambda_\alpha(t,x)\ge\delta/\|\mu\|_\infty
\big\}\ .
$$
Introducing $m_\delta:=
\min_{[\delta/\|\mu\|_\infty,\infty)}{D}>0$ we deduce from
(\ref{zz14}), (\ref{zz17}), and the previous observation that
\bean \int_0^T \big\| \nabla_x\left( \M_{2,\alpha} - \delta
\right)_+^2 \big\|_2^2\, \rd t & = & 4\, \int_0^T \int_\Om \left(
\M_{2,\alpha} - \delta \right)_+^2\, \big| \nabla_x\left(
\M_{2,\alpha} - \delta
\right)_+ \big|^2\, \rd x \rd t \\
& \le & \frac{4}{m_\delta}\, \int_0^T \int_\Om \left(
\|\mu\|_\infty\,\Lambda_\alpha - \delta \right)_+^2\,
D(\Lambda_\alpha)\, \big| \nabla_x\left( \M_{2,\alpha} - \delta
\right)_+ \big|^2\, \rd x \rd t \\
& \le & c(T,\delta)\, \int_0^T \int_\Om D(\Lambda_\alpha)\, \big|
\nabla_x\left( \M_{2,\alpha} - \delta \right)_+ \big|^2\, \rd x
\rd t \\
& \le & c(T,\delta)\ , \eean which proves the first claim in
\eqref{zz16}.

On the other hand, if $\psi\in W_{n+1}^1(\Om)$, it follows from
\eqref{zz11} and Lemma~\ref{lez1} that \bean & & \left| \int_\Om
\partial_t \left( \M_{2,\alpha} - \delta
\right)_+^2\, \psi\,\rd x \right| \\
& = & 2\,\left| \int_\Om \left( \M_{2,\alpha} - \delta \right)_+\,
\partial_t \left( \M_{2,\alpha} - \delta \right)_+\, \psi\,\rd x
\right| \\
& \le & 2\, \int_\Om \left( \M_{2,\alpha} - \delta \right)_+\,
|\nabla_x\psi| \, D(\Lambda_\alpha)\, \left|\nabla_x\M_{2,\alpha}
\right| \,\rd x \\
& + & 2\, \int_\Om |\psi| \, D(\Lambda_\alpha)\, \big|
\nabla_x\left( \M_{2,\alpha} - \delta \right)_+ \big|^2\,\rd x \\
& + & \frac{2}{\tau}\,\|\M_{2,\alpha}\|_\infty\, \|\psi\|_1\,
\big\|
\left( \M_{2,\alpha} - \delta \right)_+ \big\|_\infty \\
& + & 2\,\left( \|\mu'\|_\infty + \|\mu\|_\infty^2
\right)\,\|\M_{1,\alpha}\|_\infty\, \|\psi\|_1\, \big\| \left(
\M_{2,\alpha} - \delta \right)_+ \big\|_\infty \\
& \le & 2\, \left\| \M_{2,\alpha} \right\|_\infty\,
\|\nabla_x\psi\|_2 \, \|D(\Lambda_\alpha)\|_\infty^{1/2}\, \left(
\int_\Om D(\Lambda_\alpha)\, \big| \nabla_x\left( \M_{2,\alpha} -
\delta
\right)_+ \big|^2\,\rd x \right)^{1/2}  \\
& + & 2\,\|\psi\|_\infty\, \int_\Om \, D(\Lambda_\alpha)\, \big|
\nabla_x\left( \M_{2,\alpha} - \delta \right)_+ \big|^2\,\rd x +
c(T)\,\|\psi\|_1 \\
& \le & c(T)\, \left( \|\nabla_x\psi\|_2 + \|\psi\|_\infty
\right)\, \left( 1 + \int_\Om \, D(\Lambda_\alpha)\, \big|
\nabla_x\left( \M_{2,\alpha} - \delta \right)_+ \big|^2\,\rd x
\right)\ . \eean Owing to the continuous embedding of
$W_{n+1}^1(\Om)$ in $L_\infty(\Om)$ we conclude that
$$
\big\| \partial_t \left( \M_{2,\alpha} - \delta \right)_+^2
\big\|_{W_{n+1}^1(\Om)'} \le c(T)\, \left( 1 + \int_\Om \,
D(\Lambda_\alpha)\, \big| \nabla_x\left( \M_{2,\alpha} - \delta
\right)_+ \big|^2\,\rd x \right)\ ,
$$
which together with \eqref{zz17} implies the second claim in
\eqref{zz16}.
\end{proof}

\medskip

Lemma~\ref{lez3} provides the desired compactness for
$(\M_{2,\alpha})_\alpha$ with the help of the following lemma.

\begin{lem}\label{lez4}
Let $Q$ be an open bounded subset of $\R^N$ for some $N\ge 1$ and
$p\in [1,\infty)$. We consider a sequence $(z_k)_{k\ge 1}$ of
non-negative functions in $L_p(Q)$ and assume that there is a
sequence $(Z_j)_{j\ge 1}$ in $L_p(Q)$ such that \bqn \label{zz19}
\lim_{k\to\infty} \left\| \left( z_k - \frac{1}{j} \right)_+ - Z_j
\right\|_p =  0 \;\;\mbox{ for all }\;\; j\ge 1\ . \eqn Then
$(z_k)$ converges in $L_p(Q)$ as $k\to\infty$.
\end{lem}

\begin{proof}
For $i\ge 1$, $j\ge 1$, and $k\ge 1$ we have \bean \|Z_i-Z_j\|_p &
\le & \left\| \left( z_k - \frac{1}{i} \right)_+ - Z_i \right\|_p
+ \left\| \left( z_k - \frac{1}{i} \right)_+ - \left( z_k -
\frac{1}{j} \right)_+ \right\|_p + \left\| \left( z_k -
\frac{1}{j}
\right)_+ - Z_j \right\|_p \\
& \le & \left\| \left( z_k - \frac{1}{i} \right)_+ - Z_i
\right\|_p + |Q|^{1/p}\,\left| \frac{1}{i} - \frac{1}{j} \right| +
\left\| \left( z_k - \frac{1}{j} \right)_+ - Z_j \right\|_p\ .
\eean Letting $k\to\infty$ and using \eqref{zz19} give
$$
\|Z_i-Z_j\|_p \le |Q|^{1/p}\,\left| \frac{1}{i} - \frac{1}{j}
\right|\ ,
$$
so that $(Z_j)$ is a Cauchy sequence in $L_p(Q)$ and there is
$Z\in L_p(Q)$ such that \bqn \label{zz20} \lim_{j\to\infty} \| Z_j
- Z \|_p = 0\ . \eqn Next, for $j\ge 1$ and $k\ge 1$, we have
\bean \|z_k-Z\|_p & \le & \left\| z_k - \left( z_k - \frac{1}{j}
\right)_+ \right\|_p + \left\| \left( z_k - \frac{1}{j} \right)_+
- Z_j
\right\|_p + \left\| Z_j - Z \right\|_p \\
& \le & \frac{|Q|}{j} + \left\| \left( z_k - \frac{1}{j} \right)_+
- Z_j \right\|_p + \left\| Z_j - Z \right\|_p \ , \eean hence
$$
\limsup_{k\to\infty} \|z_k-Z\|_p \le \frac{|Q|}{j} + \left\| Z_j -
Z \right\|_p
$$
by \eqref{zz19}. Letting $j\to\infty$ and using \eqref{zz20} give
the expected convergence.
\end{proof}

\medskip

Finally, to link the limits of $(\Lambda_\alpha)_\alpha$,
$(\M_{1,\alpha})_\alpha$ and $(\M_{2,\alpha})_\alpha$ with that of
$(u_\alpha)_\alpha$ we need to control the behavior of $u_\alpha$
for large $a$ and report the following result in that direction.

\begin{lem}\label{lez6}
For $\alpha\in (0,1)$, $t\in [0,T]$, and $A\ge 1$ we have
$$
\int_\Om \int_A^\infty e^{a/\tau}\, u_\alpha(t,a,x)\,\rd a\rd x
\le c_5(T)\,\omega_\alpha(A) \;\;\mbox{ with }\;\;
\omega_\alpha(A) := \int_\Om \int_{A/2}^\infty
e^{a/\tau}\,u_\alpha^0(a,x)\,\rd a\rd x + \frac{1}{A}\ .
$$
\end{lem}

\begin{proof}
Let $\eta\in C^\infty(\R)$ be a fixed non-decreasing function such
that $\eta(a)=0$ for $a\le 1/2$ and $\eta(a)=1$ for $a\ge 1$. For
$A\ge 1$, we multiply \eqref{zz3} by $\eta(a/A)\,e^{a/\tau}$ and
integrate over $(0,\infty)\times\Om$ with the help of \eqref{zz7}.
Since $\eta(0)=0$ we thus obtain
$$
\frac{d}{dt} \int_\Om \int_0^\infty \eta\left( \frac{a}{A}
\right)\,e^{a/\tau}\,u_\alpha(t,a,x)\,\rd a\rd x \le \int_\Om
\int_0^\infty \left[ \frac{1}{A}\,\partial_a\eta\left( \frac{a}{A}
\right) + \frac{1}{\tau}\,\eta\left( \frac{a}{A} \right)
\right]\,e^{a/\tau}\,u_\alpha(t,a,x)\,\rd a\rd x\ ,
$$
$$
\frac{d}{dt} \left( e^{-t/\tau}\,\int_\Om \int_0^\infty \eta\left(
\frac{a}{A} \right)\,e^{a/\tau}\,u_\alpha(t,a,x)\,\rd a\rd x
\right) \le \frac{\|\eta\|_{W_\infty^1}}{A}\,\int_\Om
\M_{1,\alpha}(t,x)\,\rd x\ .
$$
By virtue of \eqref{zz12} the right-hand side of the above
differential inequality is bounded by $c(T)/A$ and
Lemma~\ref{lez6} follows after time integration, taking into
account the properties of $\eta$.
\end{proof}

\medskip

\noindent\textit{Proof of Theorem~\ref{wsdgn}.} Recall that $\big(
\widehat{D}(\Lambda_\alpha) \big)_\alpha$ is bounded in
$L_\infty((0,T),W_2^1(\Om))$ and $\big( \partial_t
\widehat{D}(\Lambda_\alpha) \big)_\alpha$ is bounded in
$L_2((0,T)\times\Om)$ by \eqref{zz15}. Owing to the compactness of
the embedding of $W_2^1(\Om)$ in $L_2(\Om)$ we may apply
\cite[Corollary~4]{Si87} to conclude that \bqn \label{zz24}
\big(\widehat{D}(\Lambda_\alpha) \big)_\alpha \;\;\mbox{ is
relatively compact in }\;\; C([0,T],L_2(\Om))\ . \eqn A similar
argument allows us to deduce from Lemma~\ref{lez3} and
\cite[Corollary~4]{Si87} that \mbox{$\big(\big( \M_{2,\alpha} -
1/j \big)_+^2 \big)_\alpha$} is relatively compact in
$L_2((0,T)\times\Om)$ for each $j\ge 1$. Since $\left(
\M_{2,\alpha} \right)_\alpha$ is bounded in
$L_\infty((0,T)\times\Om)$ by \eqref{zz12}, the Lebesgue dominated
convergence theorem actually allows us to conclude that \bqn
\label{zz25} \big( \big( \M_{2,\alpha} - 1/j \big)_+ \big)_\alpha
\;\;\mbox{ is relatively compact in }\;\; L_ 2((0,T)\times\Om)
\eqn for each $j\ge 1$. We then infer from Lemma~\ref{lez1},
\eqref{zz14}, and \eqref{zz24} that there are a sequence
$(\alpha_k)_{k\ge 1}$, $\alpha_k\to 0$, three functions $\ell\in
C([0,T],L_2(\Om))\cap L_2((0,T),W_2^1(\Om))$, $d\in
L_2((0,T),W_2^1(\Om))$, and $u\in L_\infty(U)$, and a sequence
$(W_j)_{j\ge 1}$ in $L_2((0,T)\times\Om)$ such that
\begin{eqnarray}
\left( u_{\alpha_k} \right)_k \stackrel{*}{\rightharpoonup} u &
\mbox{
in } & L_\infty(U)\ , \label{zz26} \\
\big( D(\Lambda_{\alpha_k})\,,\, \widehat{D}(\Lambda_{\alpha_k})
\big)_k \rightharpoonup (d,\ell) & \mbox{ in } &
L_2((0,T),W_2^1(\Om))\ , \label{zz27} \\
\big( \widehat{D}(\Lambda_{\alpha_k}) \big)_k \longrightarrow \ell
& \mbox{ in } & C([0,T],L_2(\Om))\ ,\label{zz28}\\
\big( \big( \M_{2,\alpha_k} - 1/j \big)_+ \big)_k \longrightarrow
W_j & \mbox{ in } & L_2((0,T)\times\Om)\ . \nonumber
\end{eqnarray}
Combining the last convergence and Lemma~\ref{lez4} actually give
that there is $W\in L_2((0,T)\times\Om)$ such that \bqn
\label{zz29} \left( \M_{2,\alpha_k} \right)_k \longrightarrow W
\;\;\mbox{ in }\;\; L_2((0,T)\times\Om)\ . \eqn In addition, as
the function $\widehat{D}$ is a diffeomorphism from $(0,\infty)$
onto its range with inverse $\widehat{D}^{-1}$, the bound
\eqref{zz14} and the convergence \eqref{zz28} imply that \bqn
\label{zz30} \left( \Lambda_{\alpha_k} \right)_k \longrightarrow
\widehat{D}^{-1}(\ell) \;\;\mbox{ in }\;\; C([0,T],L^2(\Om))\ .
\eqn

We now claim that, if $\chi$ is a non-negative measurable function
such that $\chi(a)\le \Xi\,e^{a/\tau}$ for a.e. $a\ge 0$ and some
$\Xi\ge 0$, we have \bqn \label{zz31} \left( \M_{\chi,\alpha_k}
\right)_k \stackrel{*}{\rightharpoonup} \M_\chi \;\;\mbox{ in
}\;\; L_\infty((0,T)\times\Om)
\eqn
with
$$
\M_{\chi,\alpha}(t,x) \,:=\, \int_0^\infty \chi(a)\,
u_\alpha(t,a,x)\,\rd a \;\;\mbox{ and }\;\;
\M_\chi(t,x)\,:=\, \int_0^\infty \chi(a)\, u(t,a,x)\,\rd a
$$
for $(t,x)\in (0,T)\times\Om$. Indeed, consider $\psi\in
L_\infty((0,T)\times\Om)$. For $A>0$ we have \bean & & \left|
\int_0^T \int_\Om \left( \M_{\chi,\alpha_k} - \M_\chi
\right)(t,x)\, \psi(t,x)\, \rd x\rd t \right| \\
& \le & \left| \int_0^T \int_\Om \int_0^A \left( u_{\alpha_k} - u
\right)(t,a,x)\, \psi(t,x)\, \chi(a)\,\rd a\rd x\rd t \right|\\
& + & \left| \int_0^T \int_\Om \int_A^\infty \left( u_{\alpha_k} -
u
\right)(t,a,x)\, \psi(t,x)\, \chi(a)\,\rd a\rd x\rd t \right|\\
& \le & \left| \int_0^T \int_\Om \int_0^A \left( u_{\alpha_k} - u
\right)(t,a,x)\, \psi(t,x)\, \chi(a)\,\rd a\rd x\rd t \right|\\
& + & \Xi\,\|\psi\|_\infty\, \int_0^T \int_\Om \int_A^\infty
e^{a/\tau}\, \left( u_{\alpha_k} + u \right)(t,a,x) \,\rd a\rd
x\rd t\ . \eean We then infer from \eqref{zz1}, Lemma~\ref{lez6},
and \eqref{zz26} by a weak convergence argument that
$$
\int_0^T \int_\Om \int_A^\infty e^{a/\tau}\, u(t,a,x)\,\rd a\rd
x\rd t \le T\,c_5(T)\,\omega_0(A) \;\;\mbox{ with }\;\;
\omega_0(A) := \int_\Om \int_{A/2}^\infty
e^{a/\tau}\,u^0(a,x)\,\rd a\rd x + \frac{1}{A}\ .
$$
It also follows from \eqref{zz1} and Lemma~\ref{lez6} that
$$
\int_0^T \int_\Om \int_A^\infty e^{a/\tau}\, u_{\alpha_k}(t,a,x)
\,\rd a\rd x\rd t \le T\,c_5(T)\, \left( \int_\Om \int_0^\infty
e^{a/\tau}\,\left| u_{\alpha_k}^0 - u^0 \right|\,\rd a\rd x +
\omega_0(A) \right)\ ,
$$
and thus
$$
\limsup_{k\to\infty} \int_0^T \int_\Om \int_A^\infty e^{a/\tau}\,
u_{\alpha_k}(t,a,x) \,\rd a\rd x\rd t \le T\,c_5(T)\,\omega_0(A)\
.
$$
Consequently, by \eqref{zz26},
$$
\limsup_{k\to\infty} \left| \int_0^T \int_\Om \left(
\M_{\chi,\alpha_k} - \M_\chi \right)(t,x)\, \psi(t,x)\, \rd x\rd t
\right| \le c(T)\,\Xi\,\|\psi\|_\infty\,\omega_0(A)\ .
$$
Since the above inequality holds true for all $A>0$ and
$\omega_0(A)\to 0$ as $A\to\infty$ by \eqref{zhid1}, we may let
$A\to\infty$ to conclude that $\left( \M_{\chi,\alpha_k}
\right)_k$ converges weakly towards $\M_\chi$ in
$L_1((0,T)\times\Om)$ as $k\to\infty$. As
$0\,\le\,\M_{\chi,\alpha_k}\,\le\,\Xi\,\M_{1,\alpha_k}$, $\left(
\M_{1,\alpha_k} \right)_k$ is bounded in
$L_\infty((0,T)\times\Om)$ by \eqref{zz12}, and since
$(0,T)\times\Omega$ has finite measure, the previous $L_1$-weak
convergence implies the claim \eqref{zz31}.

\medskip

In particular, we deduce from \eqref{zz29} (with
$\chi(a)=\mathbf{1}_{[a_0,\infty)}(a)\,e^{a/\tau}$ and
$\chi(a)=e^{a/\tau}\,\mu(a)$, respectively) that $\left(
\Lambda_{\alpha_k} \right)_k$ and $\left( \M_{2,\alpha_k}
\right)_k$ converge weakly-$*$ towards $\Lambda$ and $\M_2$ in
$L_\infty((0,T)\times\Om)$, respectively, with $\Lambda$ and
$\M_2$ given by
$$
\Lambda(t,x) := \int_{a_0}^\infty e^{a/\tau}\,u(t,a,x)\,\rd a
\;\;\mbox{ and }\;\; \M_2(t,x) := \int_0^\infty
e^{a/\tau}\,\mu(a)\,u(t,a,x)\,\rd a
$$
for $(t,x)\in (0,T)\times\Om$. Combining this fact with
\eqref{zz29} and \eqref{zz30} leads us to the identities
$\Lambda=\widehat{D}^{-1}(\ell)$ and $\M_2=W$, and we have thus
shown that \bqn \label{zz32} \left( \Lambda_{\alpha_k} \right)_k
\longrightarrow \Lambda \;\;\mbox{ in }\;\; C([0,T],L_2(\Om))
\;\;\mbox{ and }\;\; \left( \M_{2,\alpha_k} \right)_k
\longrightarrow \M_2 \;\;\mbox{ in }\;\; L_2((0,T)\times\Om)\ .
\eqn A simple consequence of \eqref{zz27} and \eqref{zz32} is that
$d=D(\Lambda)$ so that \bqn \label{zz33} D(\Lambda)\in
L_2((0,T),W_2^1(\Om)) \;\;\mbox{ and }\;\; \left(
D(\Lambda_{\alpha_k}) \right)_k \rightharpoonup D(\Lambda) \mbox{
in } L_2((0,T),W_2^1(\Om))\ . \eqn

\medskip

We next denote by $v$ the unique solution to \bqn \label{zz34}
\partial_t v(t,x) = \frac{1}{\tau}\,\big( 1 - \xi(v(t,x)) \big)\,
v(t,x) + \M_2(t,x)\ , \quad (t,x)\in (0,T)\times\Om\ , \eqn with
initial condition $v(0)=v^0$. At this stage it is rather easy to
deduce from \eqref{zz4}, \eqref{zz34}, and the properties of $\xi$
that
$$
\frac{d}{dt} \| v_{\alpha_k} - v\|_2^2 \le c_6(T)\, \big( \|
v_{\alpha_k} - v\|_2^2 + \| \M_{2,\alpha_k} - \M_2\|_2^2\big)\ .
$$
The strong convergence \eqref{zz32} in $L_2((0,T)\times\Om)$ of
$\left( \M_{2,\alpha_k} \right)_k$ towards $\M_2$, \eqref{zz1},
\eqref{zz2}, and the above differential inequality imply that \bqn
\label{zz35} \left( v_{\alpha_k} \right)_k \longrightarrow v
\;\;\mbox{ in }\;\; C([0,T],L_2(\Om))\ . \eqn

\medskip

Introducing
$$
\mathbf{J}_\alpha\,:=\, D(\Lambda_\alpha)\,\nabla_x u_\alpha
\;\;\mbox{ and }\;\; \mathbf{j}_\alpha\,:=\,
D(\Lambda_\alpha)^{1/2}\,\nabla_x u_\alpha\ ,
$$
we infer from \eqref{zz3}, \eqref{zz6}, \eqref{zz7}, and
\eqref{zz8} that
\begin{eqnarray}
0 & = & \int_0^T \int_\Om \int_0^\infty u_\alpha(t,a,x)\, \big(
\partial_t\psi\,+\,\partial_a \psi \big)(t,a,x)\,\rd a\rd
x\rd t + \int_\Om \int_0^\infty u_\alpha^0(a,x)\,\psi(0,a,x)\,\rd
a\rd
x \nonumber\\
& + & \frac{1}{\tau}\, \int_0^T \int_\Om \xi(v_\alpha(t,x))\,
v_\alpha(t,x)\, \psi(t,0,x)\,\rd x\rd t - \int_0^T \int_\Om \int_0^\infty
\mu(a)\,u_\alpha(t,a,x)\,\psi(t,a,x)\,\rd a\rd x\rd t \nonumber \\
& - & \int_0^T \int_\Om
\int_0^\infty \mathbf{J}_\alpha(t,a,x)\, \nabla_x \psi(t,a,x)\,\rd
a\rd x\rd t  \label{zz35a}
\end{eqnarray}
for $\psi\in C_c^1([0,T)\times [0,\infty)\times\bar{\Omega})$
satisfying $\partial_\nu \psi(t,a,x)=0$ for $(t,a,x)\in
(0,T)\times (0,\infty)\times\partial\Om$. By \eqref{zz12a}
$(\mathbf{j}_\alpha)_\alpha$ is bounded in $L_2(U,\R^n)$ and so is
$\mathbf{J}_\alpha=D(\Lambda_\alpha)^{1/2}\,\mathbf{j}_\alpha$ by
\eqref{zz14}. We may then assume (after possibly extracting a
further subsequence) that there is $\mathbf{J}\in L_2(U,\R^n)$
such that \bqn \label{zz36} \left( \mathbf{J}_{\alpha_k} \right)_k
\rightharpoonup \mathbf{J} \;\;\mbox{ in }\;\; L_2(U,\R^n)\ . \eqn
Owing to \eqref{zz1}, \eqref{zz26}, \eqref{zz35}, and
\eqref{zz36}, we may pass to the limit as $k\to\infty$ in the weak
formulation \eqref{zz35a} of \eqref{zz3} to conclude that \bean 0 & = & \int_0^T \int_\Om \int_0^\infty u(t,a,x)\, \big(
\partial_t\psi\,+\,\partial_a \psi \big)(t,a,x)\,\rd a\rd
x\rd t + \int_\Om \int_0^\infty u^0(a,x)\,\psi(0,a,x)\,\rd a\rd x \\
& + & \frac{1}{\tau}\, \int_0^T \int_\Om \xi(v(t,x))\, v(t,x)\,
\psi(t,0,x)\,\rd x\rd t - \int_0^T \int_\Om \int_0^\infty
\mu(a)\,\psi(t,a,x)\,\rd a\rd x\rd t \\
& - &  \int_0^T \int_\Om \int_0^\infty
\mathbf{J}(t,a,x)\, \nabla_x \psi(t,a,x)\,\rd a\rd x\rd t \eean
for $\psi\in C_c^1([0,T)\times [0,\infty)\times\bar{\Omega})$
satisfying $\partial_\nu \psi(t,a,x)=0$ for $(t,a,x)\in
(0,T)\times (0,\infty)\times\partial\Om$ as claimed in
Theorem~\ref{wsdgn}.

It remains to identify $\mathbf{J}$: for that purpose we introduce
the sets
$$
\mathcal{P}\,:=\, \big\{ (t,x)\in(0,T)\times\Om\ : \
\Lambda(t,x)>0 \big\}\ , \quad \mathcal{Z}\,:=\, \big\{
(t,x)\in(0,T)\times\Om\ : \ \Lambda(t,x)=0 \big\}\ ,
$$
and observe that $\mathbf{J}_\alpha$ may be written \bqn
\label{zz+1} \mathbf{J}_\alpha = \nabla_x \left( u_\alpha\,
D(\Lambda_\alpha) \right) - u_\alpha\,\nabla_x D(\Lambda_\alpha)
\;\;\mbox{ in }\;\; \mathcal{D}'(U,\R^n)\ . \eqn It follows at once
from \eqref{zz14}, \eqref{zz26}, \eqref{zz32}, and the continuity
of $D$ that \bqn \label{zz+2} \left(
u_{\alpha_k}\,D(\Lambda_{\alpha_k}) \right)_k \rightharpoonup
u\,D(\Lambda) \;\;\mbox{ in }\;\; L_2(U)\ . \eqn Next, we claim
that, after possibly extracting a further subsequence (not
relabeled), we have \bqn \label{zz+3} \big(
\nabla_x\widehat{D}(\Lambda_{\alpha_k}) \big)_k \longrightarrow
\nabla_x\widehat{D}(\Lambda) \;\;\mbox{ in }\;\; L_2(U,\R^n)
\;\;\mbox{ and a.e. in }\;\; (0,T)\times\Om\ , \eqn and adapt the
proof of \cite[Eq.~(3.22)]{AMST99} to this end. We multiply
\eqref{zz9} by $\widehat{D}(\Lambda_\alpha) -
\widehat{D}(\Lambda)$ and integrate over $(0,T)\times\Om$ to
obtain \bean \int_0^T \int_\Om \big( \widehat{D}(\Lambda_\alpha) -
\widehat{D}(\Lambda) \big)\, \partial_t\Lambda_\alpha\,\rd x\rd t
& = & - \int_0^T \int_\Om \nabla_x \big(
\widehat{D}(\Lambda_\alpha) - \widehat{D}(\Lambda) \big)\,\cdot\,
\nabla_x
\widehat{D}(\Lambda_\alpha)\, \rd x\rd t \\
& + & \int_0^T \int_\Om \big( g_\alpha^1 - g_\alpha^2 \big)\,
\big( \widehat{D}(\Lambda_\alpha) - \widehat{D}(\Lambda) \big)\,
\rd x\rd t\ , \eean hence
\begin{eqnarray}
\int_0^T \int_\Om \left| \nabla_x \big(
\widehat{D}(\Lambda_\alpha) - \widehat{D}(\Lambda) \big)
\right|^2\, \rd x\rd t &
= & - \int_0^T \int_\Om \nabla_x \big( \widehat{D}(\Lambda_\alpha) - \widehat{D}(\Lambda) \big)\,\cdot\, \nabla_x \widehat{D}(\Lambda)\, \rd x\rd t \nonumber\\
& + & \int_0^T \int_\Om \big( g_\alpha^1 - g_\alpha^2 \big)\, \big( \widehat{D}(\Lambda_\alpha) - \widehat{D}(\Lambda) \big)\,  \rd x\rd t \nonumber\\
& + & \int_0^T \int_\Om \big( \widehat{D}(\Lambda) -
\widehat{D}(\Lambda_\alpha) \big)\,
\partial_t\Lambda_\alpha\,\rd x\rd t\ . \label{zz+4}
\end{eqnarray}
As $\nabla_x\widehat{D}(\Lambda)\in L_2((0,T)\times\Om)$ by
\eqref{zz27} and $\widehat{D}(\Lambda)=\ell$ by \eqref{zz30} and
\eqref{zz32} we infer from \eqref{zz27} that
$$
\lim_{k\to\infty} \int_0^T \int_\Om \nabla_x \big(
\widehat{D}(\Lambda_{\alpha_k}) - \widehat{D}(\Lambda)
\big)\,\cdot\, \nabla_x \widehat{D}(\Lambda)\, \rd x\rd t = 0\ .
$$
It next follows from \eqref{zz12} and \eqref{zz14} that
$\|g_\alpha^1\|_\infty + \|g_\alpha^2\|_\infty\le c(T)$.
Therefore, by virtue of \eqref{zz28}, \eqref{zz30}, and
\eqref{zz32} we have
$$
\lim_{k\to\infty} \int_0^T \int_\Om \big( g_{\alpha_k}^1 -
g_{\alpha_k}^2 \big)\, \big( \widehat{D}(\Lambda_{\alpha_k}) -
\widehat{D}(\Lambda) \big)\,  \rd x\rd t = 0\ .
$$
We next argue as in \cite{AMST99} (with $\widehat{D}$ instead of
$z\mapsto z^m$ and $X=W_2^1(\Om)$) to show that
$$
\lim_{k\to\infty} \int_0^T \int_\Om \big( \widehat{D}(\Lambda) -
\widehat{D}(\Lambda_{\alpha_k}) \big)\,
\partial_t\Lambda_{\alpha_k}\,\rd x\rd t = 0\ .
$$
Taking $\alpha=\alpha_k$ in \eqref{zz+4} we may therefore pass to
the limit as $k\to\infty$ and conclude that \eqref{zz+3} holds
true.

Now, as $\nabla_x D(\Lambda_{\alpha_k}) = \big(
D'(\Lambda_{\alpha_k})\,\nabla_x \widehat{D}(\Lambda_{\alpha_k})
\big)/ D(\Lambda_{\alpha_k})$ we deduce from \eqref{zz32} and
\eqref{zz+3} that \bqn \label{zz+5}\big( \nabla_x
D(\Lambda_{\alpha_k}) \big)_k \longrightarrow
\frac{D'(\Lambda)}{D(\Lambda)}\,\nabla_x \widehat{D}(\Lambda) =
\nabla_x D(\Lambda) \;\;\mbox{ a.e. in }\;\; \mathcal{P}\ . \eqn
Moreover, by \eqref{zz14}, \bean
\int_\mathcal{Z} \left| \nabla_x D(\Lambda_{\alpha_k}) \right|\,\rd x\rd t & = & \int_\mathcal{Z} \frac{|D'(\Lambda_{\alpha_k})|\,\Lambda_{\alpha_k}^{1/2}}{D(\Lambda_{\alpha_k})^{1/2}}\, \frac{D(\Lambda_{\alpha_k})^{1/2}}{\Lambda_{\alpha_k}^{1/2}}\,\left| \nabla_x \Lambda_{\alpha_k} \right|\,\rd x\rd t \\
& \le & \left( \int_\mathcal{Z} \frac{|D'(\Lambda_{\alpha_k})|^2\,\Lambda_{\alpha_k}}{D(\Lambda_{\alpha_k})}\,\rd x\rd t \right)^{1/2}\,\left( \int_0^T \int_\Om \frac{D(\Lambda_{\alpha_k})}{\Lambda_{\alpha_k}}\,\left| \nabla_x \Lambda_{\alpha_k} \right|^2\,\rd x\rd t \right)^{1/2} \\
& \le & c(T)\, \left( \int_\mathcal{Z}
\frac{|D'(\Lambda_{\alpha_k})|^2\,\Lambda_{\alpha_k}}{D(\Lambda_{\alpha_k})}\,\rd
x\rd t \right)^{1/2}\ , \eean and the right-hand side of the above
inequality converges to zero as $k\to\infty$ by \eqref{z0},
\eqref{zz32}, the definition of $\mathcal{Z}$, and the Lebesgue
dominated convergence theorem. As $\nabla_x D(\Lambda)=0$ a.e. in
$\mathcal{Z}$ by Stampacchia's theorem, we have shown that, after
possibly extracting a further subsequence (not relabeled), $\left(
\nabla_x D(\Lambda_{\alpha_k}) \right)_k$ converges to $\nabla_x
D(\Lambda)$ a.e. in $\mathcal{Z}$. Recalling \eqref{zz+5} we have
actually established that $\left( \nabla_x D(\Lambda_{\alpha_k})
\right)_k$ converges to $\nabla_x D(\Lambda)$ a.e. in
$(0,T)\times\Om$ which, together with \eqref{zz33} and the Vitali
theorem, implies that $\left( \nabla_x D(\Lambda_{\alpha_k})
\right)_k$ converges to $\nabla_x D(\Lambda)$ in
$L_1((0,T)\times\Om,\R^n)$. Combining this convergence with
\eqref{zz26} entails that \bqn \label{zz+6} \left( u_{\alpha_k}\,
\nabla_x D(\Lambda_{\alpha_k}) \right)_k \rightharpoonup
u\,\nabla_x D(\Lambda) \;\;\mbox{ in }\;\; L_1(U,\R^n)\,. \eqn Thanks
to \eqref{zz+1}, \eqref{zz+2}, and \eqref{zz+6}, we conclude that
$\mathbf{J}= \nabla_x( u\, D(\Lambda) ) - u\,\nabla_x D(\Lambda)$
in $\mathcal{D}'(U,\R^n)$ as claimed in \eqref{spirou}. In fact, as
$\mathbf{J}$ and $u\,\nabla_x D(\Lambda)$ both belong to
$L_2(U,\R^n)$, we realize that $u\,D(\Lambda)$ belongs to
$L_2((0,T)\times (0,\infty),W_2^1(\Om))$. \qed

\medskip

\begin{rem}
It is actually not necessary to assume that $D\in C^{2-}(\R)$ and
Theorem~\ref{wsdgn} is valid if $D\in C(\R)$, so that it applies
to the diffusivity
$$
D(\Lambda)\,= \,D_0\, \max\{\Lambda-\Lambda_{min},0\}^{m-1} +
e^{-1/(\varepsilon\,\Lambda)}
$$
for $m>1$ and $\varepsilon>0$. The proof is nevertheless slightly
more technical as the sequence $(D_\alpha)_\alpha$ approximating
$D$ cannot coincide with $D$ on some interval and has to be
constructed carefully so that the above proof still works.
\end{rem}

\section{Concluding Remarks}

A model accounting for the swarming of the bacteria {\it Proteus
mirabilis} and involving age and spatial variables has been
studied. It describes the evolution of small non-moving cells
(swimmers) and larger moving cells (swarmers), the latter moving
according to Brownian movement $\divv_x\big(D(\Lambda)\,\nabla_x
u\big)$ with a diffusivity $D(\Lambda)$ depending on the total
motile swarmer cell biomass $\Lambda$ defined in \eqref{6} and
thus, in a nonlocal way (with respect to age), on $u$. Assuming
that the diffusivity is bounded from below by a positive real
number, existence and uniqueness of a strong solution has been
established in Section~\ref{sect2}. It is, however, expected on
biological grounds that a certain amount of biomass is required
for the motion of swarmers to be initiated, that is, $D(\Lambda)$
is expected to vanish when $\Lambda$ is below a threshold value
$\Lambda_{min}\ge 0$. A step in that direction is made in
Section~\ref{sect3} where the existence of a weak solution is
obtained for $\Lambda_{min}=0$. To our knowledge, the more
realistic case $\Lambda_{min}>0$ has not been investigated so
far, and we hope to return to this problem and to the formation of
regular patterns as well in the near future.

As a final comment, let us point out that in the model studied in
this paper only Brownian motion is responsible for the movement of
swarmer cells and describes somehow local displacements. Though,
as pointed out in \cite{ES} and \cite{Ayati1}, only swarmer cells
of a certain maturity can actively participate in group migration,
the so-called ``raft building'', but nothing prevents young
swarmers from being caught up in the flow and thus move with
larger swarmers in the rafts. The diffusion term
$\mathrm{div}_x(D(\Lambda)\nabla_x u)$, however, reflects active
movement of swarmers of any age, i.e. also of young swarmers. It
is therefore more realistic to model migration by a drift term
along the gradient of biomass, namely, $\divv_x\big( u
\,E(\Lambda,v)\,\nabla_x\Lambda\big)$ with $E\ge 0$. The velocity
$E(\Lambda,v)\,\nabla_x\Lambda$ then points in the direction of
increasing biomass density. The swarmer cell density equation
including the above two spatial mechanisms then reads \bqn
\label{asuivre}
\partial_t u+\partial_a u\,=\, \divv_x
\big(D(\Lambda)\,\nabla_x u + u
\,E(\Lambda,v)\,\nabla_x\Lambda\big)\, -\,\mu(a)\, u\ , \quad
(t,a,x)\in (0,\infty)\times(0,\infty)\times\Om\ , \eqn instead of
\eqref{1}. The special case $E(\Lambda,v)=D'(\Lambda)$ is actually
stated in \cite{Ayati1}. Note that the choice $D\equiv 0$ is
possible in \eqref{asuivre} and it would be interesting to see
whether regular structures also arise from the model accounting
only for drift motion. From a more theoretical viewpoint, the
study of \eqref{asuivre} seems to be more complicated than that of
\eqref{1} because the initial-boundary value problem is no longer
diagonal. Nevertheless existence of weak solutions can still be
established and will appear elsewhere \cite{LW}.

\section*{Acknowledgement}
\noindent This paper was initiated during a stay of the second
author at the Universit\'{e} Paul Sabatier~-~Toulouse~3. He thanks
the Institute of Mathematics for this opportunity and gratefully
acknowledges the kind hospitality and support. At that time he
held a position at the Department of Mathematics at Vanderbilt
University in Nashville.

\end{document}